\newif\ifmnras
\mnrastrue
\ifmnras
	\documentclass[useAMS,usenatbib,usedcolumn]{mnras}
\else
	\documentclass[apj,numberedappendix]{emulateapj}
\fi
\usepackage{graphics,epsf}
\usepackage{amsmath}                
\usepackage{amsfonts}               
\usepackage{amssymb}                
\usepackage{epsfig}                 
\usepackage{graphicx}
\usepackage{float}
\usepackage[usenames, dvipsnames]{color}
\usepackage[para,online,flushleft]{threeparttable}

\newcommand{\kms}{{~\rm km\; s^{-1}}}

\newcommand{\cm}{{~\rm cm}}
\newcommand{\km}{{~\rm km}}
\newcommand{\s}{{~\rm s}}

\newcommand{\g}{{~\rm g}}
\newcommand{\G}{{~\rm G}}

\newcommand{\zams}{\mathrm{ZAMS}}

\ifmnras
	\title[Rotational shear in pre-collapse massive stars]{The rotational shear in pre-collapse cores of massive stars}	
	\author[N. Zilberman, A. Gilkis and N. Soker]{Noa Zilberman, Avishai Gilkis \& Noam Soker \\
	Department of Physics, Technion -- Israel, Institute of Technology, Haifa 3200003, Israel; \\
	noazilber@campus.technion.ac.il; agilkis@alumni.technion.ac.il; soker@physics.technion.ac.il}
\fi

\begin{document}

\ifmnras
	\pagerange{\pageref{firstpage}--\pageref{lastpage}} \pubyear{2017}
	\maketitle
	
	\label{firstpage}
\else
	\title{The rotational shear in pre-collapse cores of massive stars}

	\author{Noa Zilberman, Avishai Gilkis \& Noam Soker\altaffilmark{1}}

	\altaffiltext{1}{Department of Physics, Technion -- Israel Institute of Technology, Haifa
32000, Israel; noazilber@campus.technion.ac.il; agilkis@alumni.technion.ac.il; soker@physics.technion.ac.il}

\fi


\begin{abstract}
We evolve stellar models to study the rotational profiles of the pre-explosion cores of single massive stars that are progenitors of core collapse supernovae (CCSNe), and find large rotational shear above the iron core that might play an important role in the jet feedback explosion mechanism by amplifying magnetic fields before and after collapse. Initial masses of $15 M_\odot$ and $30 M_\odot$ and various values of the initial rotation velocity are considered, as well as a reduced mass-loss rate along the evolution and the effect of core-envelope coupling through magnetic fields. We find that the rotation profiles just before core collapse differ between models, but share the following properties. (1) There are narrow zones of very large rotational shear adjacent to convective zones. (2) The rotation rate of the inner core is slower than required to form a Keplerian accretion disk. (3) The outer part of the core and the envelope have non-negligible specific angular momentum compared to the last stable orbit around a black hole (BH). Our results suggest the feasibility of magnetic field amplification which might aid a jet-driven explosion leaving behind a neutron star. Alternatively, if the inner core fails in exploding the star, an accretion disk from the outer parts of the core might form and lead to a jet-driven CCSN which leaves behind a BH.
\end{abstract}


\section{INTRODUCTION}
\label{sec:intro}

A growing number of observations and their analysis support the notion that jets play a crucial role in the explosion of many core collapse supernovae (CCSNe; e.g., \citealt{Wangetal2001, Maundetal2007, Lopezetal2011, Lopezetal2013, Lopezetal2014, Milisavljevic2013, Gonzalezetal2014, Marguttietal2014, Milisavljevicetal2015, FesenMilisavljevic2016, Inserraetal2016, Mauerhanetal2017, GrichenerSoker2017, BearSoker2017, Marguttietal2017a, Piranetal2017, Tanakaetal2017}). Support to the notion comes also from numerical simulations and analytical arguments (e.g. \citealt{Khokhlovetal1999, MacFadyen2001, Hoflich2001, Woosley2005, Burrows2007, Couch2009, Couch2011, TakiwakiKotake2011, Lazzati2012, Maedaetal2012, Mostaetal2014, Nishimura2015, BrombergTchekhovskoy2016, Gilkis2017, Nishimuraetal2017, Sobacchietal2017}).
Most of these theoretical studies hold that for the newly formed neutron star (NS) or black hole (BH) to launch jets, the pre-collapse core must spin rapidly. Such rapid rotation seems to require a stellar binary companion that enters the envelope of the CCSN progenitor or that the star never reaches a giant phase.
According to most of these studies, therefore, jets take place in rare circumstances, and they attribute some roles to jets only in specific types of CCSNe. 

In contrast, there is a recent suggestion that \textit{all CCSNe are exploded by jets}, and that the jets operate in a negative feedback mechanism (e.g., \citealt{PapishSoker2011, Papishetal2015, GilkisSoker2015, Gilkisetal2016}; review by \citealt{Soker2016Rev}). In this negative feedback cycle, when the jets manage to efficiently remove mass from the vicinity of their origin, the mass accretion rate on to the central object decreases, and hence the power of the jets decreases. In the case of CCSNe, when the jets managed to eject (explode) the entire core, they shut themselves off.
 
In cases of a slowly rotating pre-collapse core, the angular momentum available for the formation of an accretion disk or belt results from pre-collapse turbulence zones in the core regions. The turbulence supplies the seeds to perturbations \citep{GilkisSoker2014, GilkisSoker2015, GilkisSoker2016} that are amplified by instabilities after collapse, mainly by the spiral modes of the standing accretion shock instability (SASI). 
The spiral modes of the SASI have been studied in recent years (e.g., \citealt{Fernandez2010, Kotakeetal2011, Rantsiouetal2011, Endeveetal2012, GuiletFernandez2014, Iwakamietal2014, Abdikamalovetal2015, Fernandez2015, Jankaetal2016, MorenoMendezCantiello2016, Kazeronietal2016, Blondinetal2017, Kazeronietal2017}), and were found, among other things, to influence the final angular momentum of the NS. \cite{Endeveetal2012} find that the spiral modes can amplify turbulence and magnetic fields. 

Rotational shear can further amplify magnetic fields in the core, before and after collapse. \cite{Wheeleretal2015} discuss the role of shear in the pre-collapse core. In particular, they examine the magnetorotational instability and its role in slowing down the core rotation. They evolve massive stars using the numerical code \textsc{mesa} (Modules for Experiments in Stellar Astrophysics; \citealt{Paxton2011,Paxton2013,Paxton2015}) and study the rotational profile in the core. \cite{Wheeleretal2015} also mention the amplification of magnetic fields as a result of the shear. Their analysis of the magnetorotational instability (MRI) in their rotating model suggests that magnetic fields of $\approx 10^{12} \G$ might exist at the edge of the iron core.

\cite{Mostaetal2015} perform three-dimensional magnetohydrodynamic simulations of collapsing rapidly rotating cores. They show that the turbulence amplifies the magnetic field and leads to a large-scale magnetic field, that might form bipolar magnetorotationally driven outflows. The angular velocity gradient above the iron core of the pre-collapse core, taken as in \cite{TakiwakiKotake2011}, is shallower than we find in the present study for a given rotation velocity. 
In any case, the general conclusion from the studies of \cite{Wheeleretal2015} and \cite{Mostaetal2015} is that the shear in the pre-collapse core can amplify magnetic fields before the collapse. \cite{Mostaetal2015} further find that amplification also takes place after the formation of the NS at the centre.

For slowly rotating cores of massive stars, the pre-collapse perturbations and post-collapse instabilities cannot bring the temporary specific angular momentum to high enough values that are needed to form an accretion disk. Namely, the specific angular momentum stays below the Keplerian one on the surface of the newly-born NS. A temporary accretion belt is formed perpendicular to the temporary direction of the angular momentum of the accreted gas.
An accretion belt is defined here to be a thick sub-Keplerian rotating accretion inflow that does not extend much beyond the NS (or BH). In addition, the specific angular momentum of the accreted gas is large enough to prevent an inflow along the two opposite polar directions. \cite{SchreierSoker2016} suggest that such an accretion belt substantially amplifies magnetic fields, and then can launch jets.

In the present paper we simulate the rotation of pre-collapse cores and obtain the rotation profile and the rotation shears just before collapse. In section \ref{sec:setup} we describe the numerical method, and in section \ref{sec:rotation} we present the pre-collapse rotation curves of the different evolutionary models. Several studies have followed the evolution of rotating stars (e.g., \citealt{Heger2000,Hegeretal2005,HirschiMeynetMaeder2004,HirschiMeynetMaeder2005,Yoon2005,WoosleyHeger2005,Yoon2006}). Our new addition is the study of the influence of the mass loss rate on the rotational profile of the core, and in comparing the angular momentum in the core to that requires to form an accretion disk around the newly born NS of BH.

In section \ref{sec:shear} we present the rotational shear for many of the models that we have simulated. 
Other papers, e.g., \cite{Hegeretal2005} and \cite{Wheeleretal2015}, have studied the shear in pre-collapse cores. We differ from them by including a study with different mass loss prescriptions and by discussing the effects of the strong shear on the jet feedback explosion mechanism. We also discuss the role that rotational shear can play in further amplifying the magnetic fields and the implications of strong magnetic fields (such as suggested by \citealt{Wheeleretal2015}) on the explosion mechanism. We examine the rotation profile and identify the location of the zones with strong rotational shear in the pre-collapse core under a variety of initial conditions and various mass-loss rates and core-envelope magnetic coupling efficiencies.
 Our summary is in section \ref{sec:summary}. 

\section{NUMERICAL SETUP AND PHYSICAL INGREDIENTS}
\label{sec:setup}

We construct a set of stellar models using Modules for Experiments in Stellar Astrophysics (\textsc{mesa} version 8845; \citealt{Paxton2011,Paxton2013,Paxton2015}), with initial masses of $M_\zams=15 M_\odot$ and $M_\zams=30 M_\odot$, and initial rotation in the range $0.05 \le \Omega \le 0.9$, where we define $\Omega \equiv \left(\omega / \omega_\mathrm{crit}\right)_\zams$. The critical angular velocity $\omega_\mathrm{crit}$ is
\begin{equation}
\omega_\mathrm{crit}^2 = \left(1-L/L_\mathrm{Edd}\right)GM/R^3,
\label{eq:wcrit}
\end{equation}
where $M$ is the total mass, $R$ is the photospheric radius\footnote{More accurately, this should be the {\it equatorial radius at critical rotation}. This is one of several inaccuracies arising in the one-dimensional treatment of rapidly-rotating stars.}, $L$ is the luminosity, and $L_\mathrm{Edd}$ is the Eddington luminosity of the star (the role of the Eddington factor is discussed in detail by \citealt{Maeder2000}).

Rotation is implemented in {\textsc{mesa} using the `shellular approximation' \citep{Meynet1997}, where the angular velocity $\omega$ is assumed to be constant over isobars. Rotationally-induced instabilities and convection transport angular momentum within the stellar models \citep{Paxton2013}. Convection is treated according to the Mixing-Length Theory with $\alpha_\mathrm{MLT}=1.5$. Semiconvective mixing \citep{Langer1983,Langer1991} is employed with $\alpha_\mathrm{sc}=0.01$. Exponential convective overshooting is applied as in \cite{Herwig2000}, with $f=0.016$ (the fraction of the pressure scale height for the decay scale). All models have an initial metallicity of $Z=0.02$, and are evolved until the onset of core-collapse (in-fall velocity of $1000 \kms$).

Our models account for angular momentum transfer by the Spruit-Tayler (ST) dynamo \citep{Spruit2002} as well. Similarly to \cite{Yoon2006}, we modify the diffusive viscosity due to magnetic torques, so that it is $\nu_\mathrm{mag} = f_{\nu, \rm{mag}} \nu_\mathrm{ST}$, where $\nu_\mathrm{ST}$ is the magnetic viscosity according to \cite{Spruit2002}, and $f_{\nu, \rm{mag}}$ is a scaling parameter. We check eleven values of $f_{\nu, \rm{mag}}$, equally spaced logarithmically, between $f_{\nu, \rm{mag}}=0.01$ and $f_{\nu, \rm{mag}}=1$. We also point out the results of \cite{Potter2012}, who predict that massive stars ($M_\zams \ga 15 M_\odot$ for LMC metallicity) cannot sustain dynamo-driven fields.

Mass-loss during the main sequence phase is treated according to the results of \cite{Vink2001}. During the giant phase, mass-loss depends on surface luminosity and temperature according to the fit of \cite{deJager1988}. Some models lose their hydrogen envelope and reach a Wolf-Rayet (WR) phase. At this point mass-loss is according to \cite{WindWR}. The mass-loss rate is enhanced by rotation (e.g., \citealt{Heger2000,Maeder2000}) by a factor of $\left(1-\omega/\omega_\mathrm{crit}\right)^{-0.43}$.

In some models, a linear scaling factor $\eta$ for the mass-loss rate was applied for the entire evolution, with values ranging from $\eta=0.1$ up to standard (canonical) mass-loss with $\eta=1$ (the same range of mass-loss reduction was also studied by \citealt{Renzo2017} for non-rotating stars). Modeling the evolution with a reduced mass-loss rate is motivated by two reasons. (1) When the mass-loss is dominated by radiation pressure on dust, the calibrated rate might represent binary systems, as most massive stars interact with a companion during their lifetime \citep{Sana2012}, and interaction is all the more likely when the envelope is inflated and red. This argument was used also to claim for a lower mass-loss rate in isolated low mass stars \citep{SabachSoker2017}. (2) Formation of clumps in stellar winds can lead to the overestimation of the mass-loss by a factor of $3$ to $10$ \citep{Smith2014}.

A total of 76 stellar models were evolved up to the point where the iron core becomes unstable and starts to collapse. Not all models in the large parameter space (initial mass, initial rotation velocity, mass-loss rate and magnetic viscosity) were simulated. Some parameter choices resulted in convergence problems -- in particular reduced mass-loss combined with high initial rotation. The main results are presented in section \ref{sec:rotation}, and additional properties of interest are listed in the appendix.

\section{PRE-COLLAPSE DIFFERENTIAL ROTATION}
\label{sec:rotation}

Fig. \ref{fig:jm15} shows the specific angular momentum profiles for models of $M_\zams=15 M_\odot$ and an initial rotation velocity on the equatorial plane of $v_\zams=131\kms$, corresponding to $\Omega \equiv \left(\omega / \omega_\mathrm{crit}\right)_\zams=0.2$ (a rather typical velocity, e.g., \citealt{TarantulaXII,TarantulaXXI}), for different assumed mass-loss rates, scaled by $\eta$, and magnetic viscosities, scaled by $f_{\nu, \rm{mag}}$. We note that the specific angular momentum shown is the average for each shell, while on the equator the specific angular momentum is larger by a factor of $1.5$, according to the shellular approximation. It can be seen that the rotation profile is not affected much by the reduction of the wind mass-loss, with the relative specific angular momentum varying by less than an order of magnitude between models. The largest deviation is seen for $\eta=0.3$, where a less massive iron core is formed. This is an example of the notorious non-monotonicity of stellar evolution. The magnetic viscosity has a more pronounced effect on the core rotation rate, with a trend of weaker core-envelope coupling leading to higher values of the core specific angular momentum.
\begin{figure*}
   \centering
   \begin{tabular}{cc}
  \includegraphics*[scale=0.28]{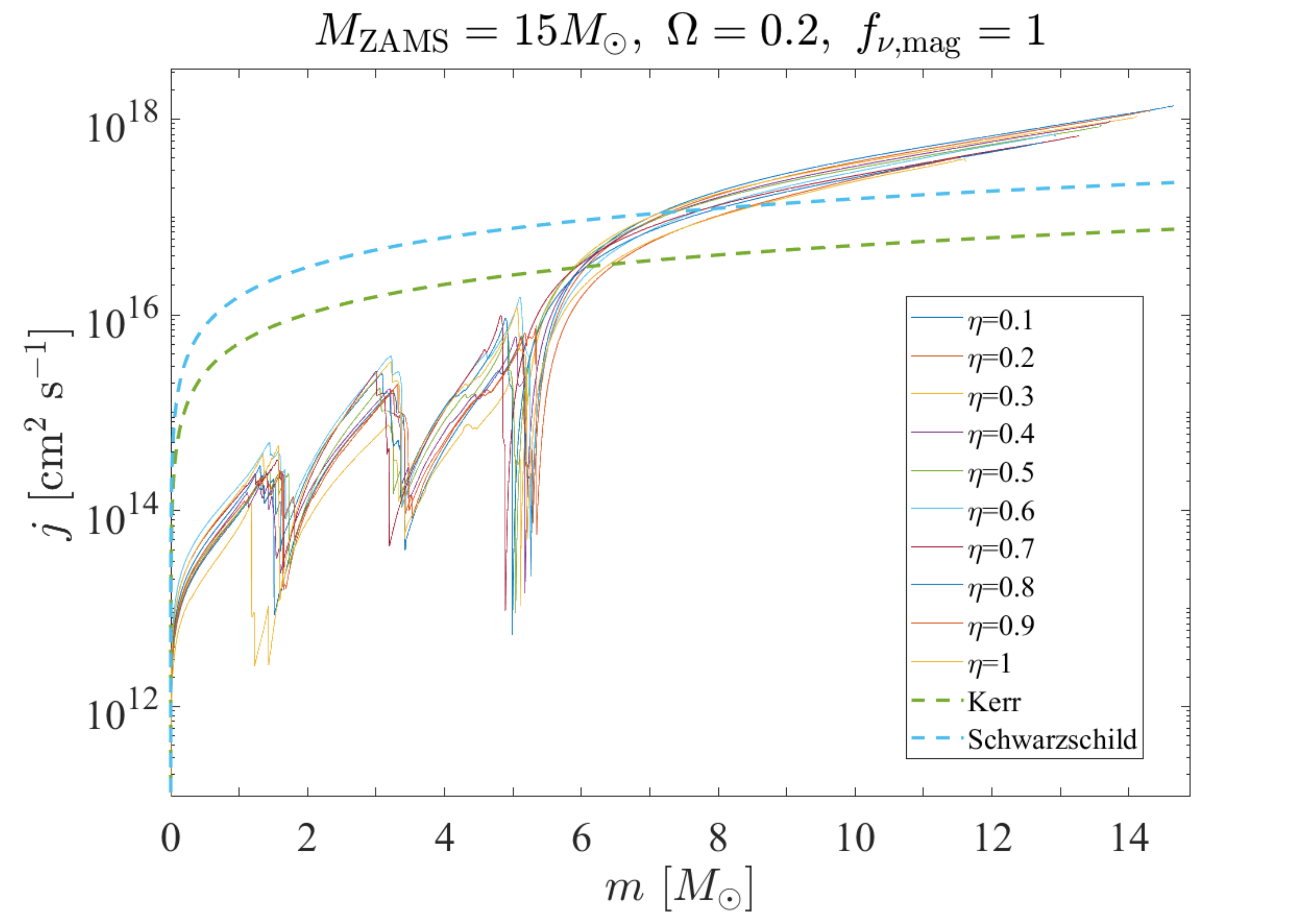} &
  \includegraphics*[scale=0.28]{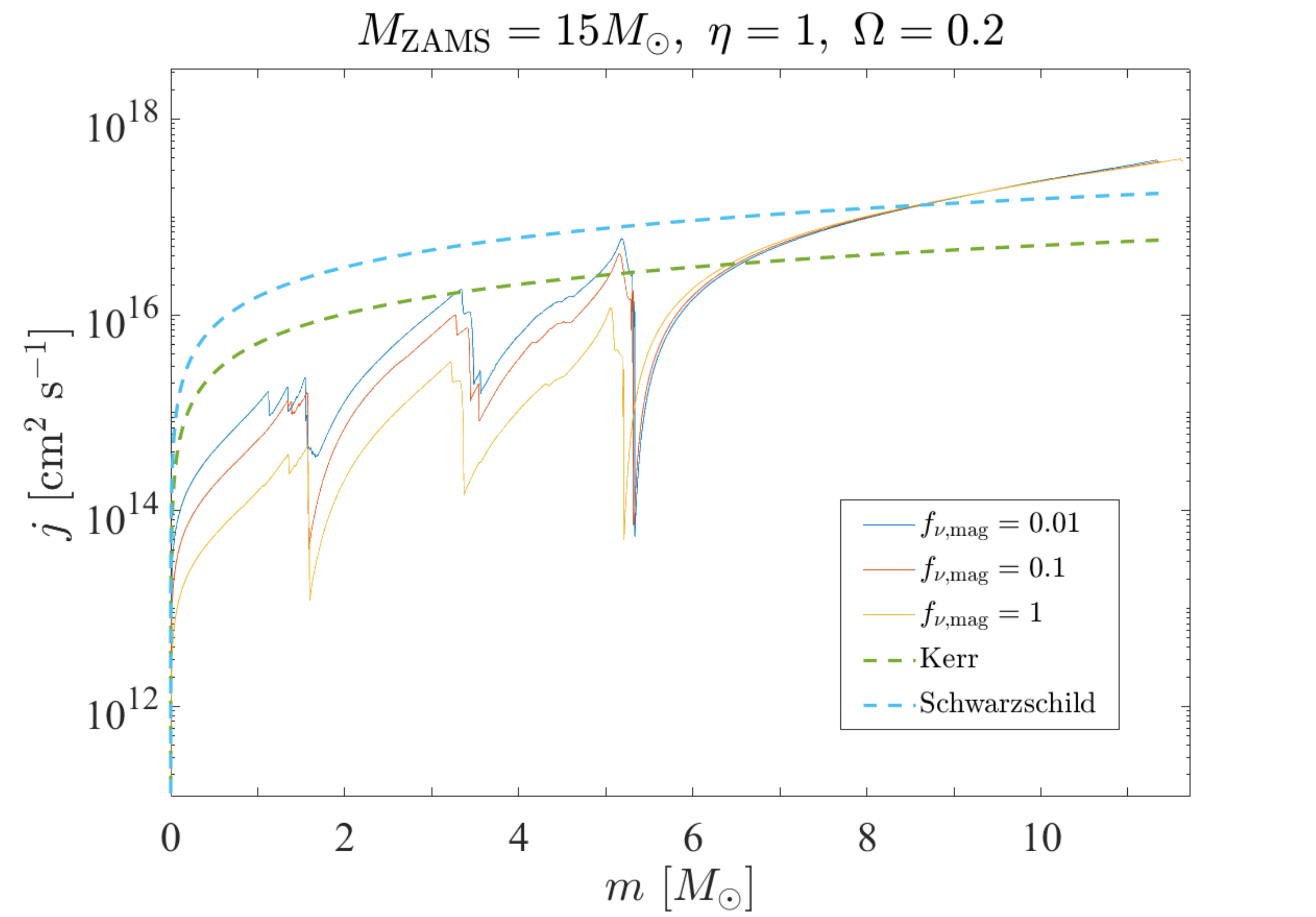} \\
  \includegraphics*[scale=0.27]{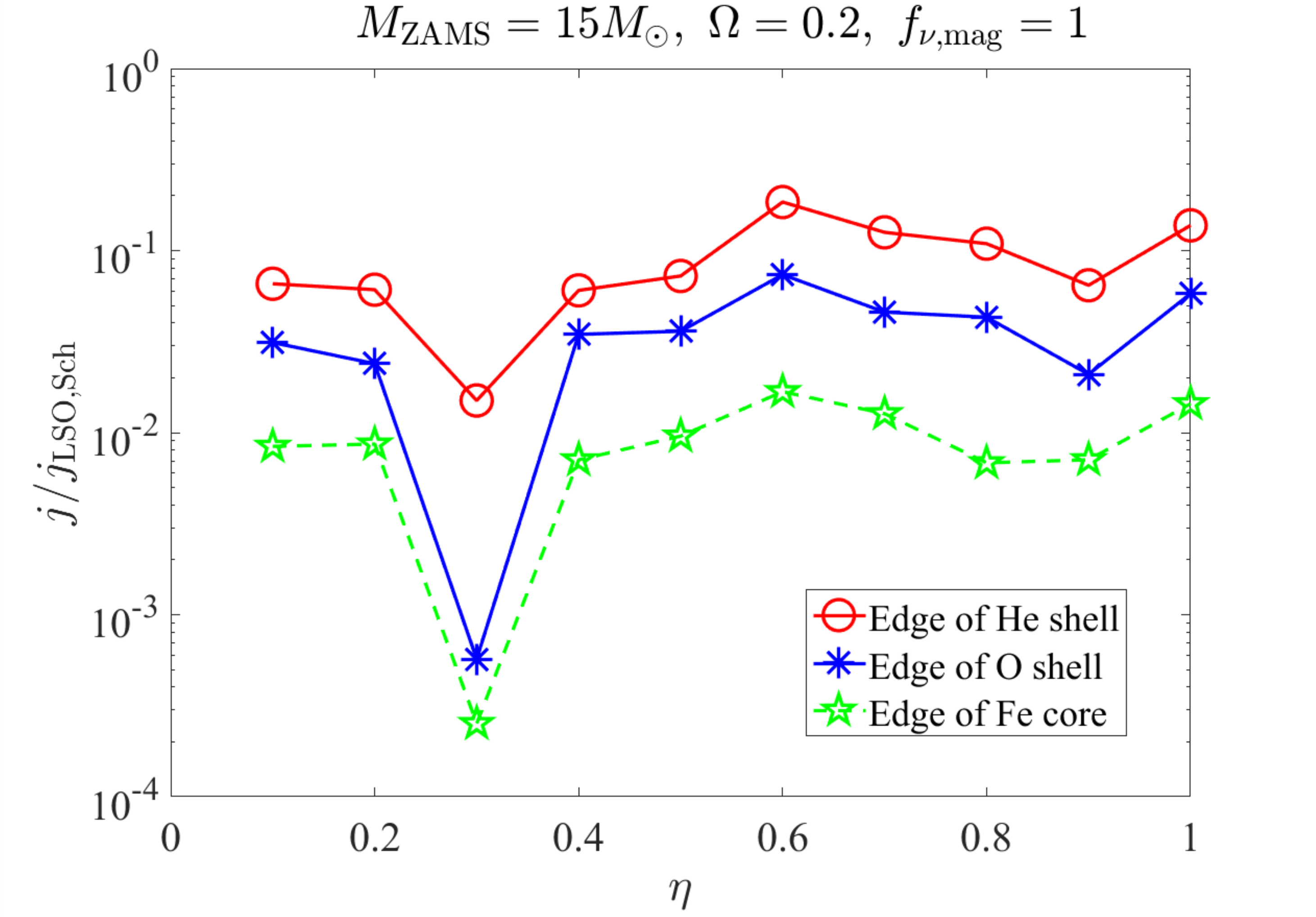} &
  \includegraphics*[scale=0.27]{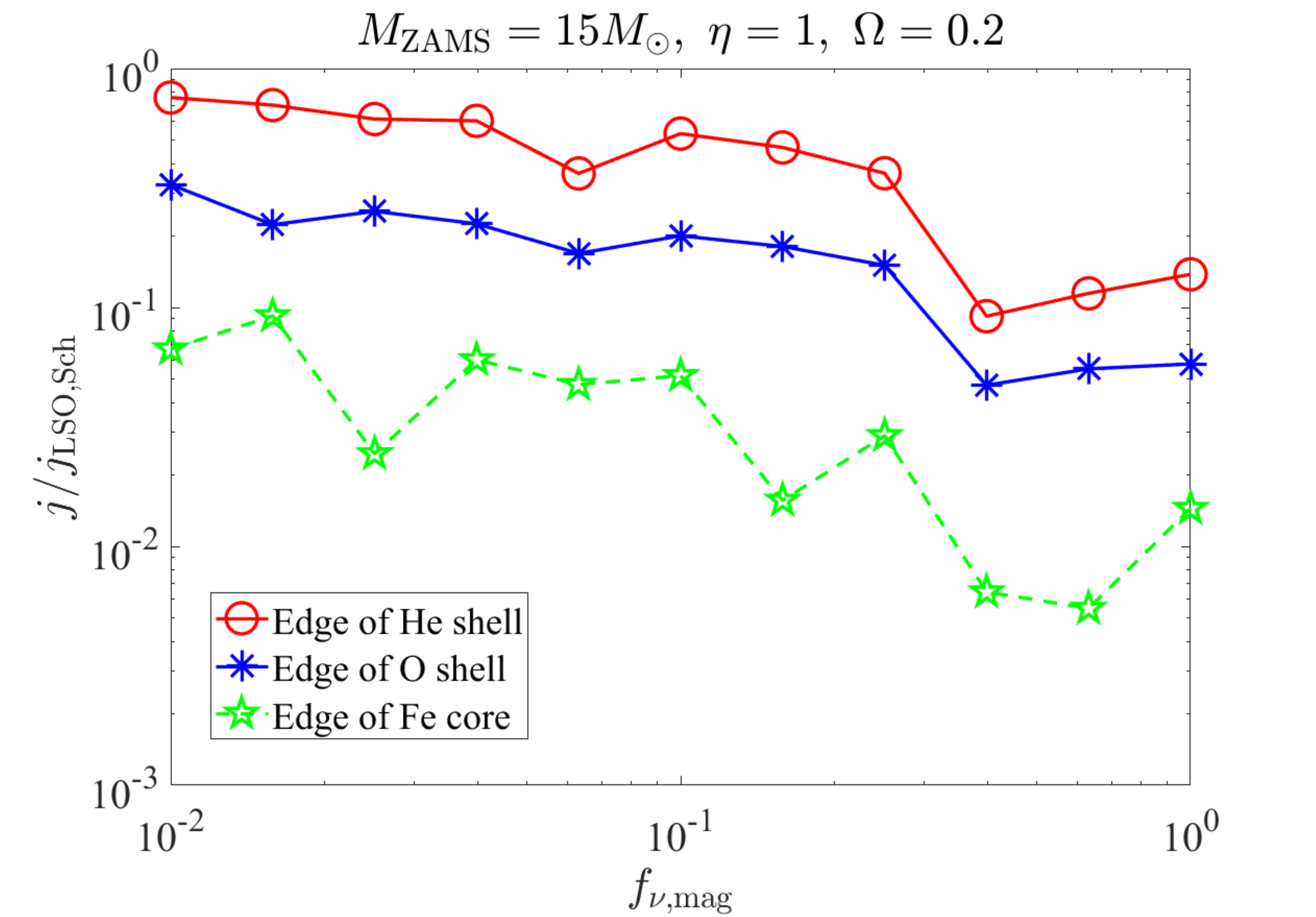} \\
  \end{tabular}
\caption{\textit{Top-left:} The specific angular momentum profiles for pre-collapse models of $M_\zams=15 M_\odot$ and $\Omega\equiv \left(\omega / \omega_\mathrm{crit}\right)_\zams=0.2$, and various mass-loss rates. For comparison, the specific angular momentum of the last stable orbit around a BH is presented, calculated at each mass coordinate for an equivalent Schwarzschild or maximally-rotating Kerr BH. The region where the specific angular momentum is above the Schwarzschild value is the hydrogen envelope. \textit{Top-right:} The specific angular momentum profiles for pre-collapse models with different magnetic viscosities. The boundaries between layers can be seen as steep changes in the specific angular momentum: the edge of the iron core at $m\approx 1.5M_\odot$; the outer oxygen layer around $m\approx 3M_\odot$; and the outer helium shell at $m\approx 5 M_\odot$. \textit{Bottom-left:} The specific angular momentum at the edges of the core layers, divided by the specific angular momentum of the innermost stable circular orbit around a Schwarzschild BH of a mass equivalent to the appropriate mass coordinate (edge of iron core, oxygen shell or helium shell), for different mass-loss rates. \textit{Bottom-right:} Relative specific angular momentum at the edges of the core layers for different magnetic viscosities.}
      \label{fig:jm15}
\end{figure*}

At each mass coordinate $m$ of the stellar models, a Schwarzschild BH with the same mass is considered, and the specific angular momentum of its innermost stable circular orbit (ISCO),
\begin{equation}
j_\mathrm{LSO,Sch} = \sqrt{12} G m / c,
\label{eq:jlso}
\end{equation}
is presented for comparison. The specific angular momentum of the core material presented in Fig. \ref{fig:jm15} is lower than that of the Schwarzschild ISCO, while for most of the hydrogen envelope it is higher. Therefore, if after collapse the core fails in driving an explosion and a BH is formed, the material of the hydrogen envelope cannot fall into the BH without forming an accretion disk. This is probably irrelevant though, as the binding energy of the hydrogen envelope is quite low, and reduction of the gravitational potential due to mass-loss by neutrino emission will generate a shock wave which ejects the envelope \citep{Nadezhin1980,Lovegrove2013}.

Fig. \ref{fig:jcore15} shows the dependence of the final core rotation rate on the initial rotation rate, for two values of the mass-loss scaling factor, $\eta=0.1$ and $\eta=1$. For canonical mass-loss, a nearly-constant relation is obtained between the initial and final rotation rates, except for several cases of large specific angular momentum in the helium shell which are to be discussed below. Another distinct feature is the separation between the shell boundaries, where for each model the relative specific angular momentum is higher in the helium shell than in the oxygen shell, and the same relation holds between the oxygen shell and iron core. This is due to the specific angular momentum increasing more rapidly with the mass coordinate than the Schwarzschild relation does.
\begin{figure}
   \centering
   \begin{tabular}{c}
  \includegraphics*[scale=0.27]{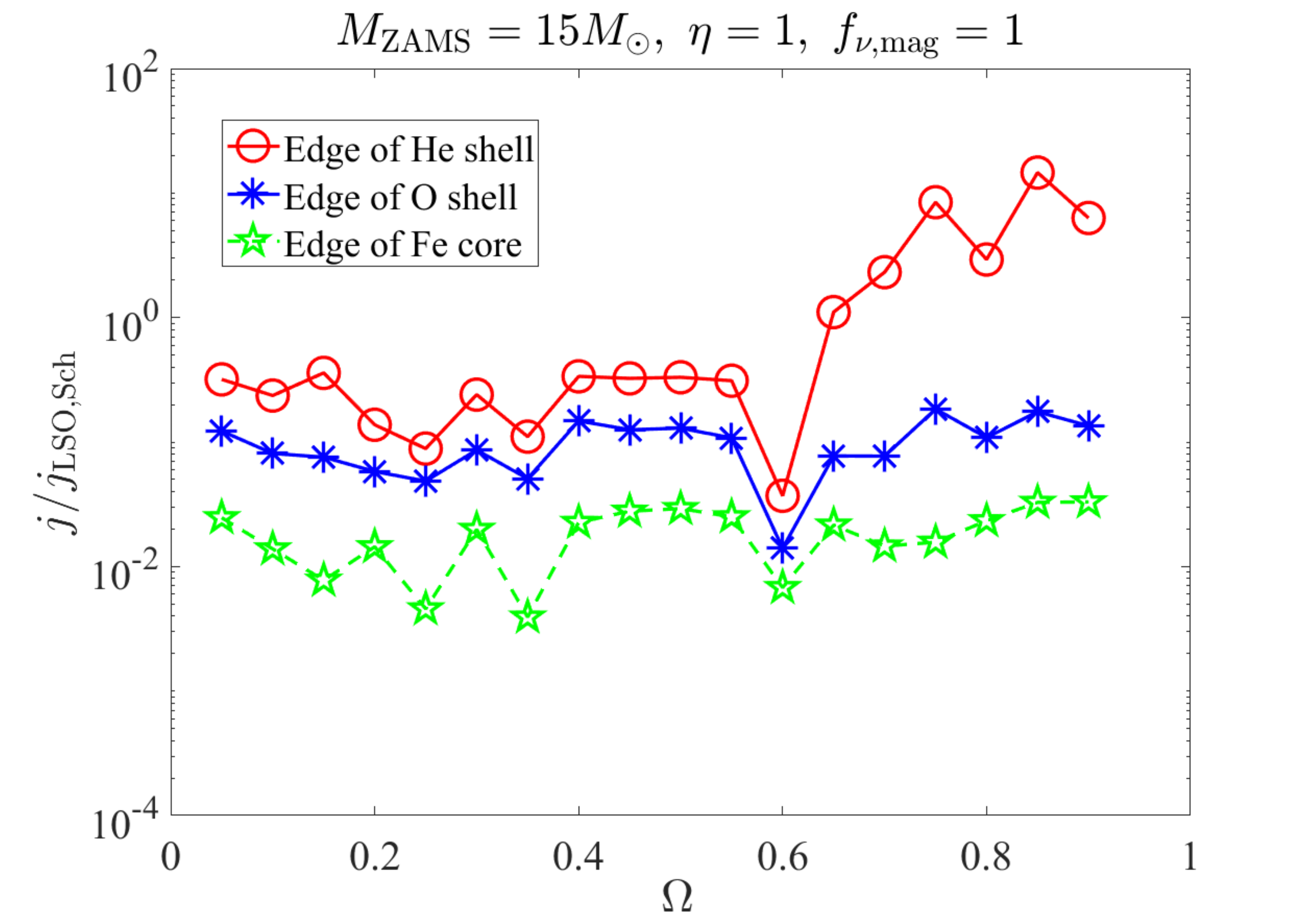} \\
  \includegraphics*[scale=0.27]{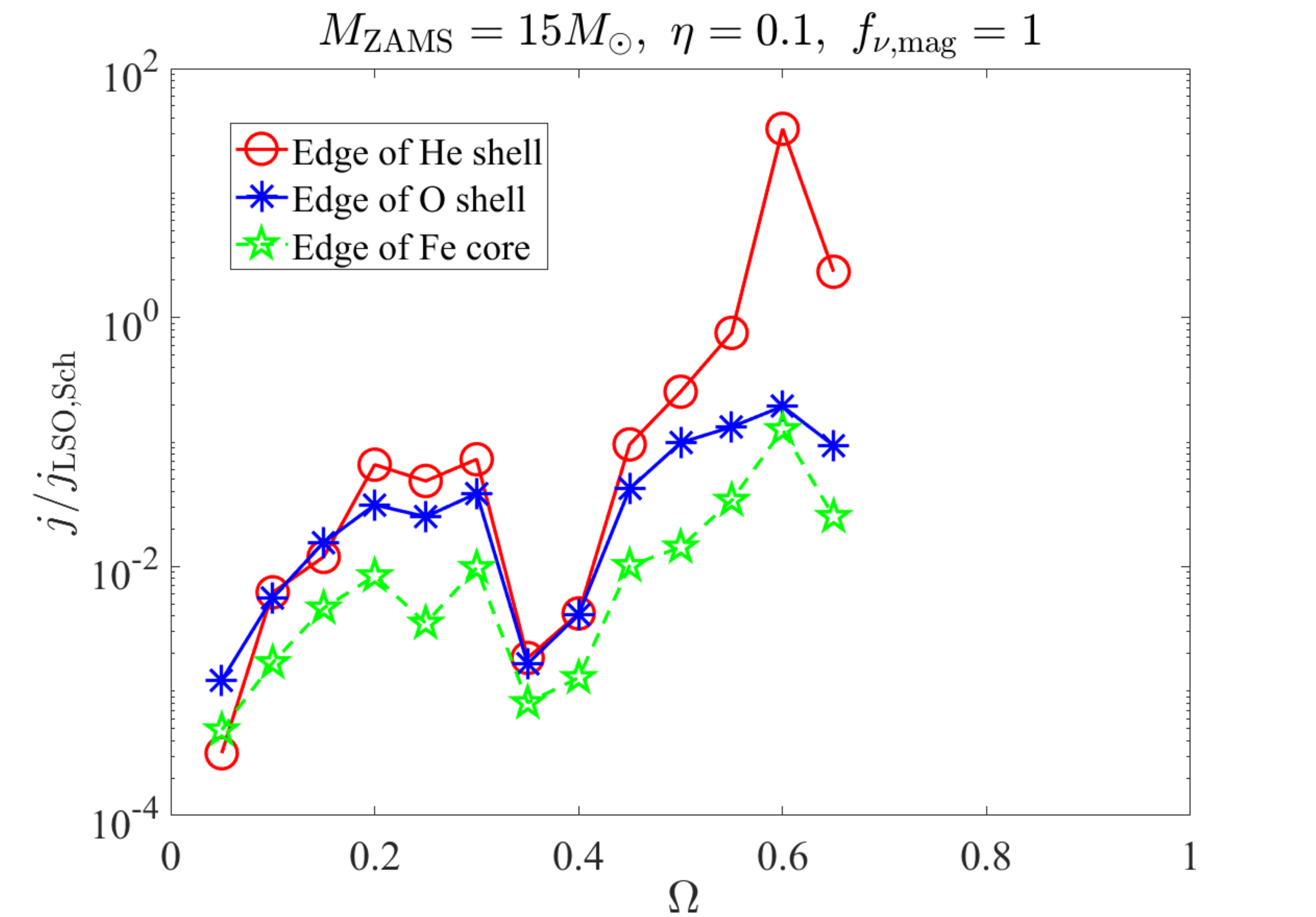} \\
  \end{tabular}
\caption{The specific angular momentum at the edge of the iron core, and at the edge of the oxygen and helium shells, for models of $M_\zams=15 M_\odot$ and different ZAMS rotation velocities. As in the bottom panels of Fig. \ref{fig:jm15}, the specific angular momentum is presented relative to the equivalent Schwarzschild value. The top panel shows the results for the standard mass-loss, and the bottom panel for reduced mass-loss by a factor of $\eta=0.1$.}
      \label{fig:jcore15}
\end{figure}

The reduced mass-loss rate yields a more pronounced correlation between the initial and final rotation rates. This is possibly due to the core-envelope coupling, with reduced mass-loss leading to a more massive hydrogen envelope, which more forcefully damps the core rotation. Although a different relation is discerned for the different mass-loss rates, for the lower initial rotation rates (which represent most massive stars) the specific angular momentum is below that of the ISCO around an equivalent mass Schwarzschild BH.

Several interesting cases are seen for very high initial rotation rates, representing very rare cases of massive stars (e.g., \citealt{TarantulaXII,TarantulaXXI}). $M_\zams=15 M_\odot$ models with $\Omega \ga 0.65$ (for $\eta=1$) and $\Omega \ga 0.55$ ($\eta=0.1$) end up with high specific angular momentum at the edge of their helium shells. All of these models end their lives as helium stars, so that the edge of the helium layer is actually the photosphere. In some cases the total mass above the point where $j=j_\mathrm{LSO,Sch}$ is negligible, but in other cases it is a few tenths of a solar mass, and in one case it even exceeds $1M_\odot$. Then, a failure of the inner core in driving a successful SN explosion will lead to the formation of a massive accretion disk ($M_\mathrm{d}\approx 0.01-1 M_\odot$) around a BH ($M_\mathrm{BH}\approx 5-10 M_\odot$), and a jet-driven CCSN.

The aforementioned models of helium stars with $j/j_\mathrm{LSO,Sch}>1$ (Fig. \ref{fig:jcore15}) reach the stage of iron core formation without ever going through a giant phase. This is similar to so-called chemically homogeneous evolution (CHE, or quasi-CHE for when the evolution is not completely chemically homogeneous), although it is usually expected to occur only for low metallicity stars (e.g., \citealt{Yoon2005,WoosleyHeger2005,Yoon2006}), and our results might arise from a shortcoming of the numerical procedure. One such point is the treatment of near critical rotation,The phenomenon of quasi-CHE in our models is questionable due to the approach of these models to critical rotation, whereupon the one-dimensional treatment of the deformed outer layers becomes uncertain. This issue is treated by limiting the angular velocity to $\omega / \omega_\mathrm{crit} \le 0.99$ through adjustment of the mass-loss rate. This is sometimes termed `mechanical mass-loss' (e.g., \citealt{Granada2013}; see also \citealt{Marchant2016}). Stellar models with high initial rotation rates contract while on the main sequence, with some of them approaching critical rotation. The models can be classified into two classes, the class evolving to the red and the class with higher initial rotation rates evolving quasi-homogeneously and remaining in the blue region of the Hertzprung-Russel diagram. Most models with $M_\zams=15M_\odot$ evolve into red supergiants (RSG), with a narrow transition around $\Omega \approx 0.5$ for which the final stage is a blue supergiant (BSG; for $\eta=0.1$) or yellow supergiant (YSG; for $\eta=1$)\footnote{We define RSG as a star with a surface temperature of $\log T_\mathrm{eff}\left(\mathrm{K}\right) < 3.6$, BSG for $\log T_\mathrm{eff}\left(\mathrm{K}\right) > 3.8$, and YSG for the cases in between.}.

Fig. \ref{fig:jcore30} shows the final core rotation rate for models with $M_\zams=30M_\odot$ and different initial rotation rates, for $\eta=0.1$ and $\eta=1$. For the models with $M_\zams=30M_\odot$ and $\eta=1$ the pre-collapse core rotation is less sensitive to the initial rotation velocity than for the models with $M_\zams=15M_\odot$. A similar feature is the low relative specific angular momentum of the core for most cases. Only for $\eta=0.1$ and $\Omega=0.5$ (corresponding to $v_\zams=329\kms$), a high pre-collapse rotation rate is reached, with the star never expanding much during its evolution. Higher initial rotation rates (for which numerical difficulties were encountered) are likely to evolve this way as well. An intriguing result is that all models with $M_\zams=30M_\odot$ and $\eta=1$ reach a giant phase, with none evolving through quasi-CHE. This might be due to the increased loss of angular momentum accompanying strong mass loss, followed by the cessation of the rotational mixing which drives homogeneous evolution.
\begin{figure}
   \centering
   \begin{tabular}{c}
  \includegraphics*[scale=0.27]{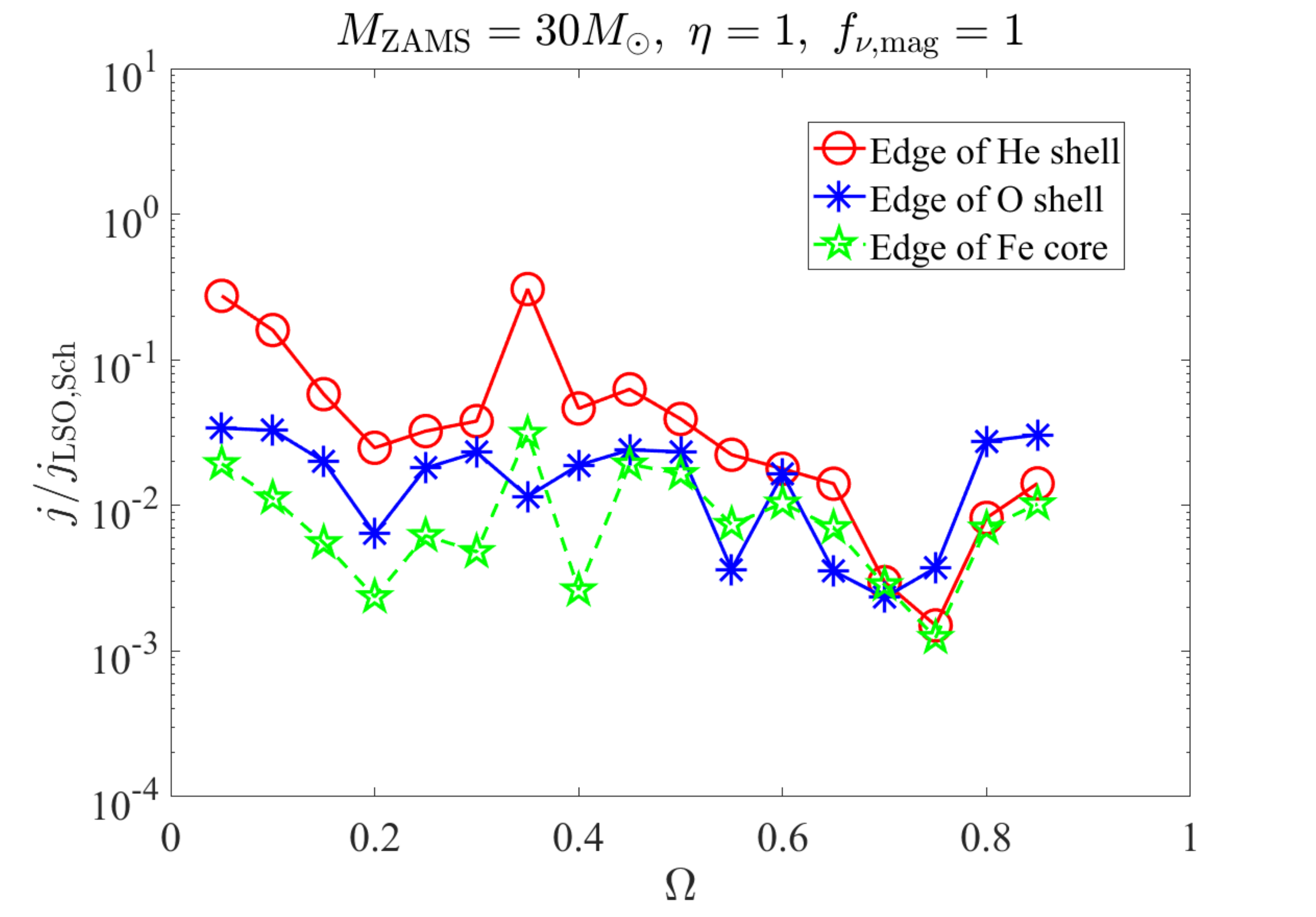} \\
  \includegraphics*[scale=0.27]{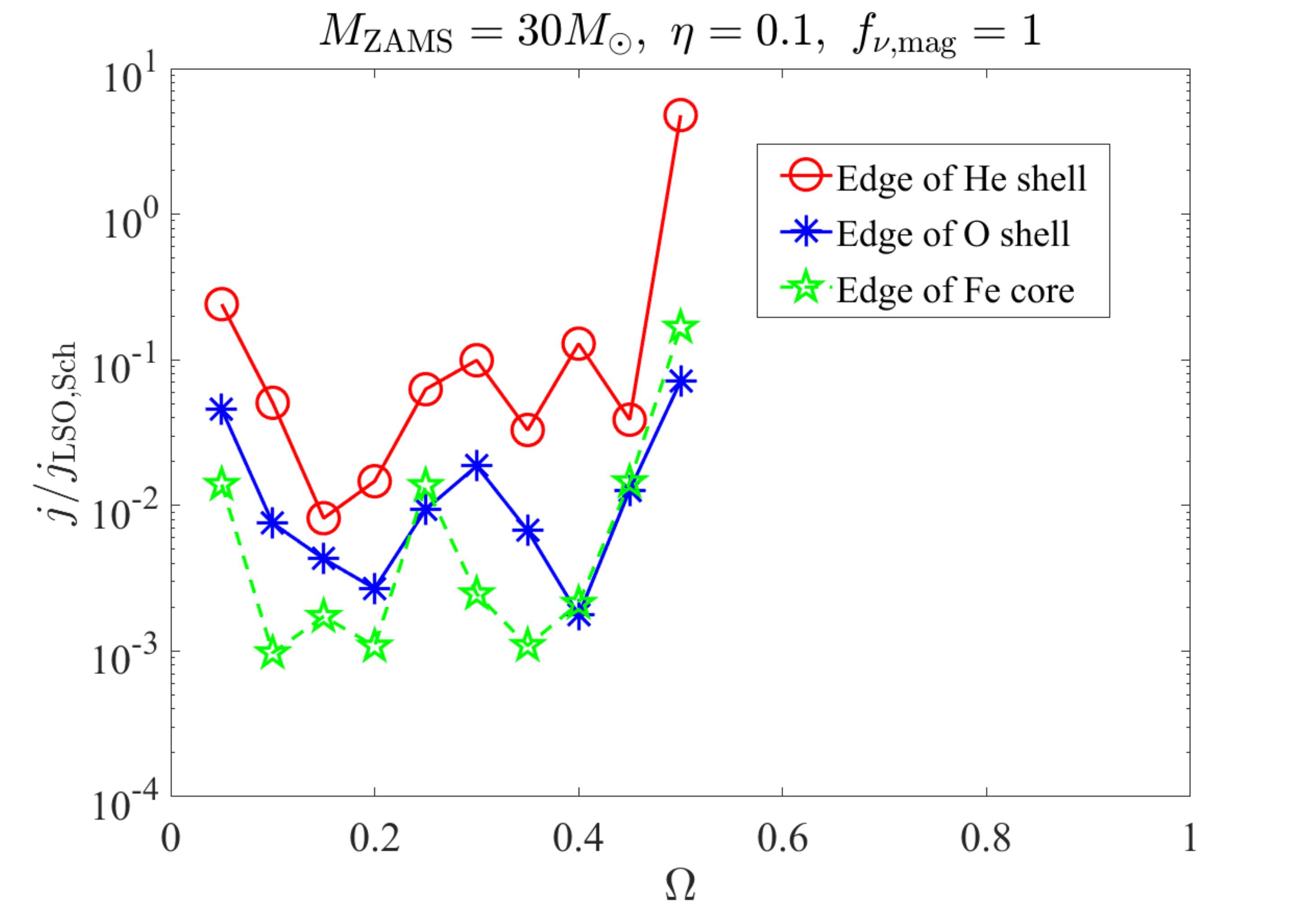} \\
  \end{tabular}
\caption{Like Fig. \ref{fig:jcore15} but for $M_\zams=30 M_\odot$.}
      \label{fig:jcore30}
\end{figure}

Most models with $M_\zams=30M_\odot$ and $\eta=1$ end their lives as WR stars\footnote{In some studies (e.g., \citealt{Renzo2017}) stellar models are classified as WR stars when the mass fraction of $^{1}\mathrm{H}$ on the surface, $X_s$, is below $0.4$. In our models we use $X_s<0.3$ for WR stars, as we find models with $0.3<X_s<0.4$ which have photospheric radii and surface temperatures typical of BSGs. It should be noted that WR stars are identified by broad emission lines, and we do not perform the required spectral modeling for this.}, while a few become BSGs with low-mass hydrogen envelopes. For $M_\zams=30M_\odot$ and $\eta=0.1$, however, a massive hydrogen envelope is retained for $\Omega < 0.45$, and the stellar models are YSGs. In most YSG models, a part of the hydrogen envelope has $j>j_\mathrm{LSO,Sch}$. As discussed by \cite{Lovegrove2013}, the effect of envelope ejection due to gravitational mass-loss by neutrino emission might be weaker for more massive stars, and accretion of the hydrogen envelope onto a BH following the failure of the core in exploding the star will produce a transient event (e.g., \citealt{WoosleyHeger2012}). In the BSG models, no part of the low-mass hydrogen envelope possesses $j>j_\mathrm{LSO,Sch}$.

To summarize this section, we find that the specific angular momentum of the inner core (up to the oxygen layer) at the pre-explosion stage is lower than required to form an accretion disk around a NS or BH after collapse. The helium layer and the hydrogen envelope, however, have non-negligible specific angular momentum in many cases. Therefore, an accretion belt or an accretion disk might form around a newly-formed BH, if the inner core fails in exploding the star. Our results do not rule out the formation of intermittent-stochastic accretion belts due to instabilities as required by the jittering-jets mechanism. In the next section we discuss the possibility of a jet-driven explosion aided by magnetic field amplification due to the rotational shear in the inner core.

\section{THE ROTATIONAL SHEAR}
\label{sec:shear}
\subsection{Numerical derivation of rotational shear}
\label{subsec:shearresults}
In this section we study the rotational shear in the pre-collapse core, and briefly discuss its possible implications. We first discuss the rotational shear above and close to the iron core, and in section \ref{subsec:blackholes} we study the shear further out in the core.
The variation of the rotational shear in the core depends on many parameters along the stellar evolution, noticeably the viscosity, in particular the magnetic coupling, e.g., the Spruit-Tayler (ST) mechanism (e.g., \citealt{Hegeretal2005}). We here study only the effects of mass-loss rate and coupling (magnetic viscosity). 

\cite{Wheeleretal2015} study the effects due to both the magnetorotational instability (MRI) and the ST mechanism. They obtain the angular velocity profiles when they include none, one of these, or both MRI and ST mechanisms. Although the angular velocity profile at the end of their calculations changes from one case to the other, in all cases there are zones of strong shear above the iron core. \cite{Wheeleretal2015} find that when the ST and MRI mechanisms are both included, the final rotation more closely resembles cases that include only the ST mechanism than those that include only the MRI. We here include (through \textsc{mesa}) the ST mechanism.  

Our analysis differs from those of \cite{Hegeretal2005} and \cite{Wheeleretal2015} in three main aspects. (1) We study the influence of reduced mass-loss rate. (2) We discuss the effects of the strong shear on the jet feedback explosion mechanism, and in particular on the jittering-jets explosion mechanism (in section \ref{subsec:shearimplicaitons}). (3) For that goal, we present the value of the shear parameter scaled by Keplerian angular velocity, 
\begin{equation}
q_{\rm Kep} = \frac {r}{\omega_{\rm Kep}} \frac {d \omega} {dr} =
 \frac {\omega}{\omega_{\rm Kep}} \frac {d \ln \omega} {d \ln r}.  
\label{eq:qkep}
\end{equation}
Here $\omega_{\rm Kep} = (GM_r/r^3)^{1/2}$ is the Keplerian angular velocity, and $\omega$ and $M_r$ are the angular velocity and mass at radius $r$ in the pre-collapse core. Since the angular velocity decreases with radius in most of the stellar interior, the value of $q_{\rm Kep}$ is generally negative. The models that we present here are the same as those studied in section \ref{sec:rotation}. 

Numerically the shear as given in equation \ref{eq:qkep} is calculated by dividing the difference in $\ln \omega$ between two adjacent numerical shells by the difference in $\ln r$ between them, and then multiply by the local value of $\omega /\omega_{\rm Kep}$. The main issue with the shear is not the numerical resolution, but rather real physical instabilities that might smear it a little. To the accuracy of our discussion to follow, the numerical procedure and resolution are adequate.

In Fig. \ref{fig:SM15eta10} we present the value of $q_{\rm Kep}$ for a model with an initial mass of $M_{\rm ZAMS} =15 M_\odot$, and for various values of the ratio of the initial angular velocity to the critical angular velocity, $\Omega$. In this figure all cases have a mass-loss scaling factor of $\eta=1$. In Fig. \ref{fig:SM15eta01} we present the shear for the same initial model, but the mass-loss rate along the entire evolution was ten times lower, i.e., $\eta=0.1$. Fig. \ref{fig:SM30eta10} and Fig. \ref{fig:SM30eta01} present the respective cases for a stellar model with an initial mass of $30 M_\odot$. In Fig. \ref{fig:SM15VarEta} we present again the model with $M_{\rm ZAMS} =15 M_\odot$, but for a value of  $\Omega = 0.2$ and different mass-loss rates. This figure shows that the details of the final rotational profile change with mass-loss rate along the evolution, but in all cases there are zones of large shear. 
\begin{figure}
      \vskip 0.5 cm
   \hskip -0.5 cm
  \includegraphics*[scale=0.59]{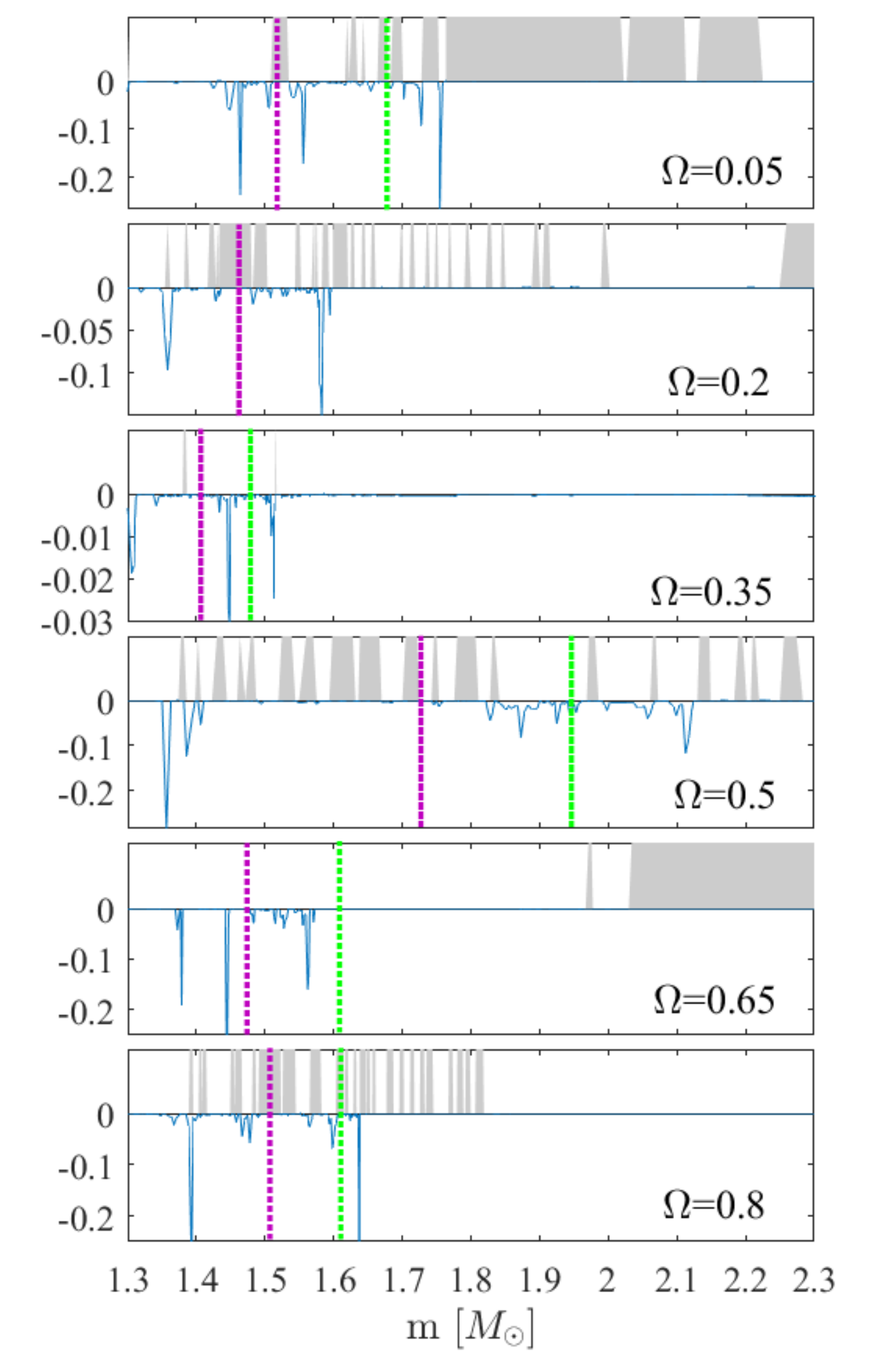} \\
\caption{The value of the shear as calculated by $q_{\rm Kep}$ that is given in equation (\ref{eq:qkep}), depicted with a blue line below the zero line, as function of the mass coordinate in the range $1.3-2.3 M_\odot$, and  for the models of $M_\zams=15 M_\odot$. The grey areas above the zero line present the convective zones in the core. The mass-loss scaling factor is $\eta=1$, and panels represent different values of the ratio of the initial angular velocity to the critical angular velocity, $\Omega$. Dashed vertical pink lines mark the edge of the Fe core, and green lines mark the edge of the Si shell.  Shears that dropped lower than $-0.2$ were cut at that minimal value. In this set, models with $\Omega=0.65$ and $0.8$ have a minimum shear of $-0.3$ and $-0.7$, respectively. In the $\Omega=0.2$ model, the Si is unmarked as silicon and oxygen are mixed above the iron core. In all cases strong shear is seen above the iron core.
}
      \label{fig:SM15eta10}
\end{figure}
\begin{figure}
      \vskip 0.5 cm
      \hskip -0.5 cm
  \includegraphics*[scale=0.585]{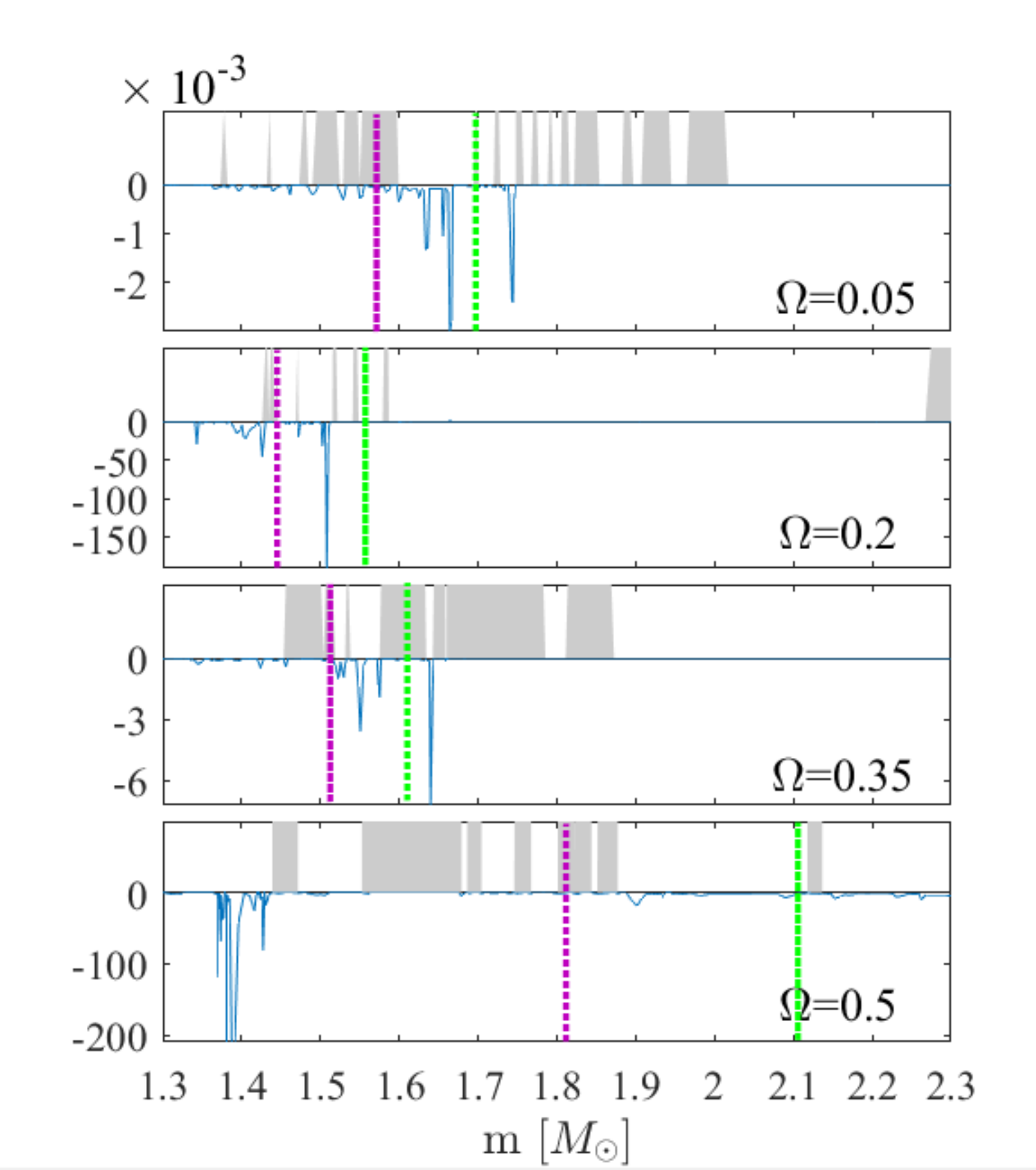} \\
\caption{Like Fig. \ref{fig:SM15eta10}, but for a mass-loss scaling factor of $\eta=0.1$, i.e., the mass-loss rate is ten times lower than that for the cases presented in Fig. \ref{fig:SM15eta10}. Note that here the marks on the vertical axis for $q_{\rm Kep}$ are in units of $10^{-3}$. In this set, models with $\Omega=0.35$ and $0.5$ had a minimum shear of $-7$ and $-11$ in that range, respectively. In all cases strong shear is seen above the iron core. }
      \label{fig:SM15eta01}
\end{figure}
\begin{figure}
      \vskip 0.5 cm
   \hskip -0.5 cm    
  \includegraphics*[scale=0.61]{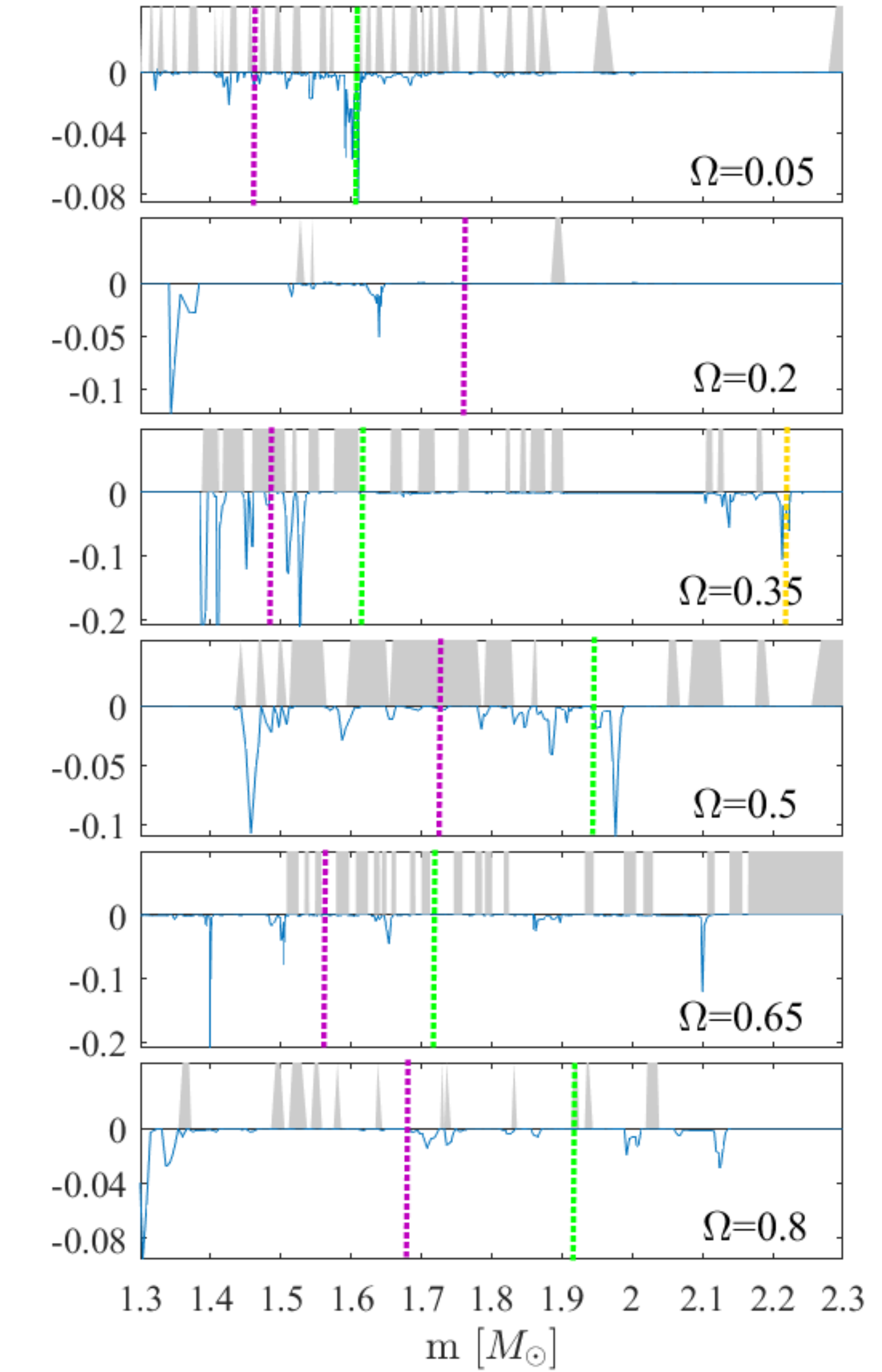} \\
\caption{Like Fig. \ref{fig:SM15eta10} but for a model with an initial mass of $M_\zams=30 M_\odot$. Pink lines mark the edge of the Fe core, and green lines mark the edge of the Si shell. In one case the edge of the O shell is within the mass range, and is marked with a yellow line. In this set, models with $\Omega=0.35$ and $0.65$ had a minimum shear of $-0.45$ and $-14$, respectively. }
      \label{fig:SM30eta10}
\end{figure}
\begin{figure}
      \vskip 0.5 cm
   \hskip -0.5 cm   
  \includegraphics*[scale=0.63]{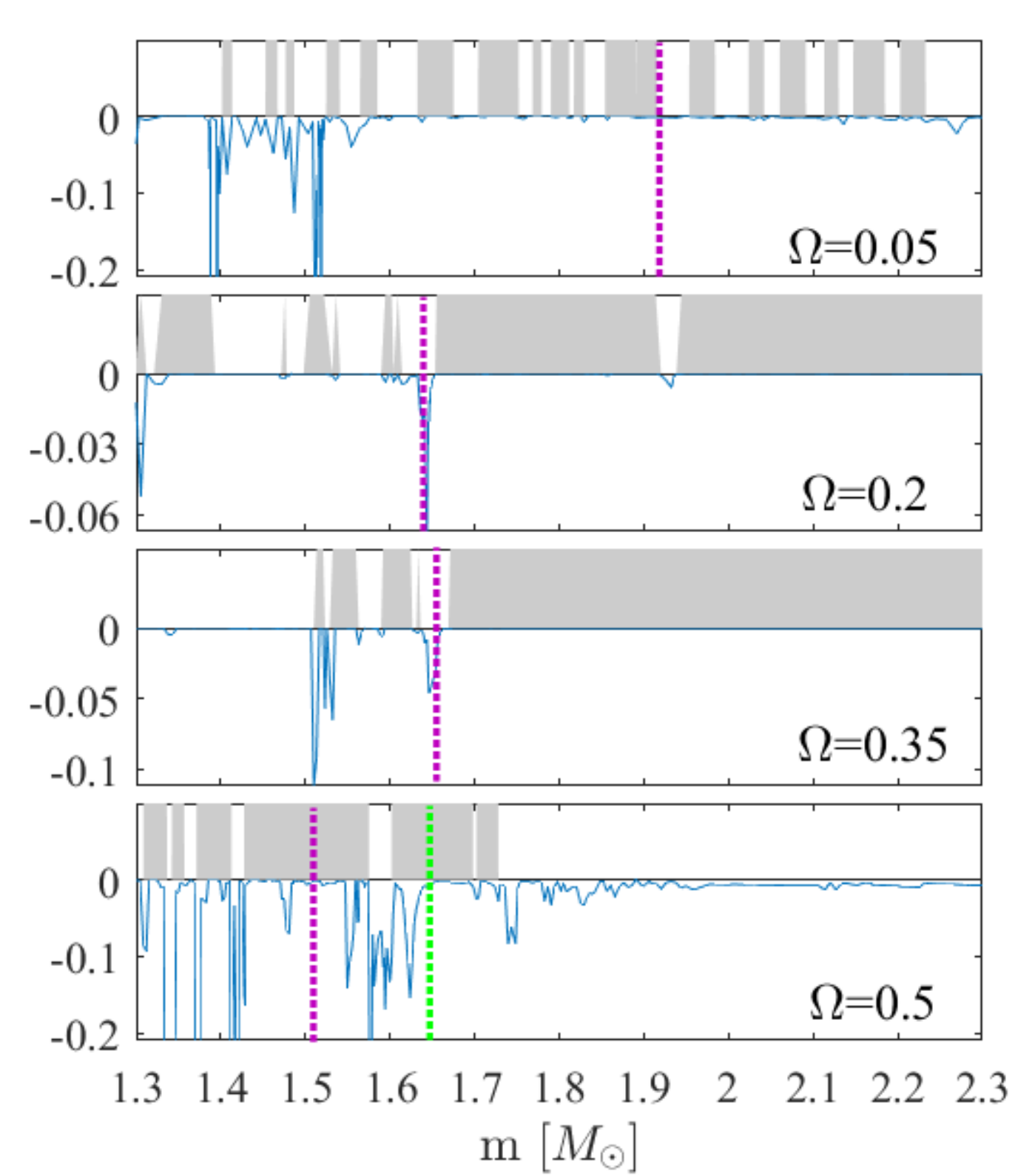} \\
\caption{Like Fig. \ref{fig:SM30eta10} but for models with a mass-loss rate ten times lower, i.e., $\eta=0.1$. In this set, models with $\Omega=0.05$ and $0.5$ had a minimum shear of $-7$ and $-4$, respectively.}
      \label{fig:SM30eta01}
\end{figure}
\begin{figure}
      \vskip 0.5 cm
      \hskip -0.5 cm 
  \includegraphics*[scale=0.62]{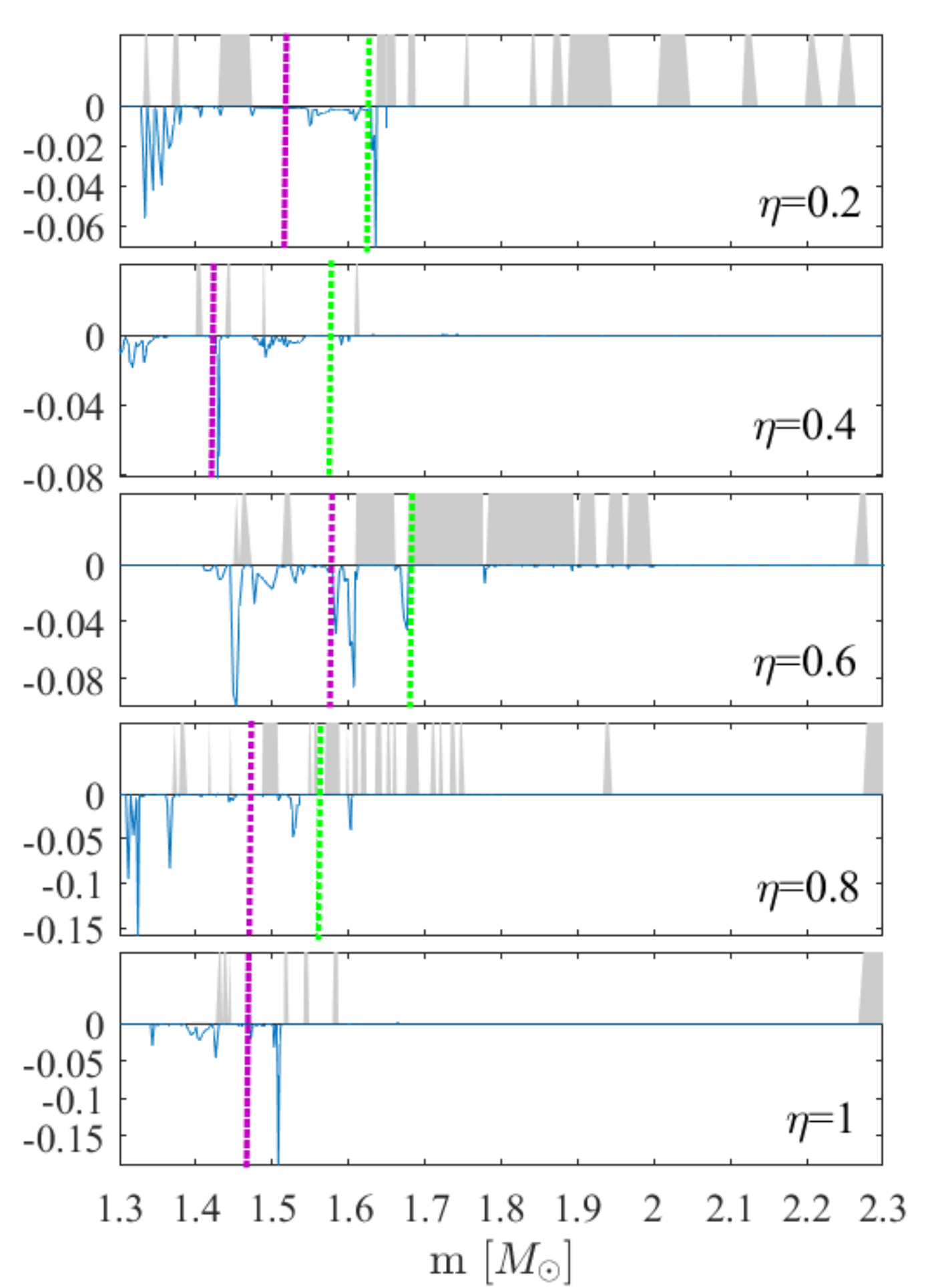} \\
\caption{Like Fig. \ref{fig:SM15eta10} but for $\Omega=0.2$ and five values of mass-loss rate parameter $\eta$. }
      \label{fig:SM15VarEta}
\end{figure}

We also examine the role of the core-envelope coupling through $f_{\nu, {\rm mag}}$ (see discussion of this parameter in section \ref{sec:rotation}). In Fig. \ref{fig:SM15varf} we present the shear as defined in equation \ref{eq:qkep}, for six cases that differ only by the value of $f_{\nu, {\rm mag}}$. The exact location of the high-shear zones and their values, and the location of the convective zones, change from one case to the other, but in all cases there are high-shear zones adjacent to convective zones.  
\begin{figure}
 \centering
      \vskip 0.5 cm
    \hskip -0.5 cm 
  \includegraphics*[scale=0.77, trim={1cm 0 17cm 0},clip]{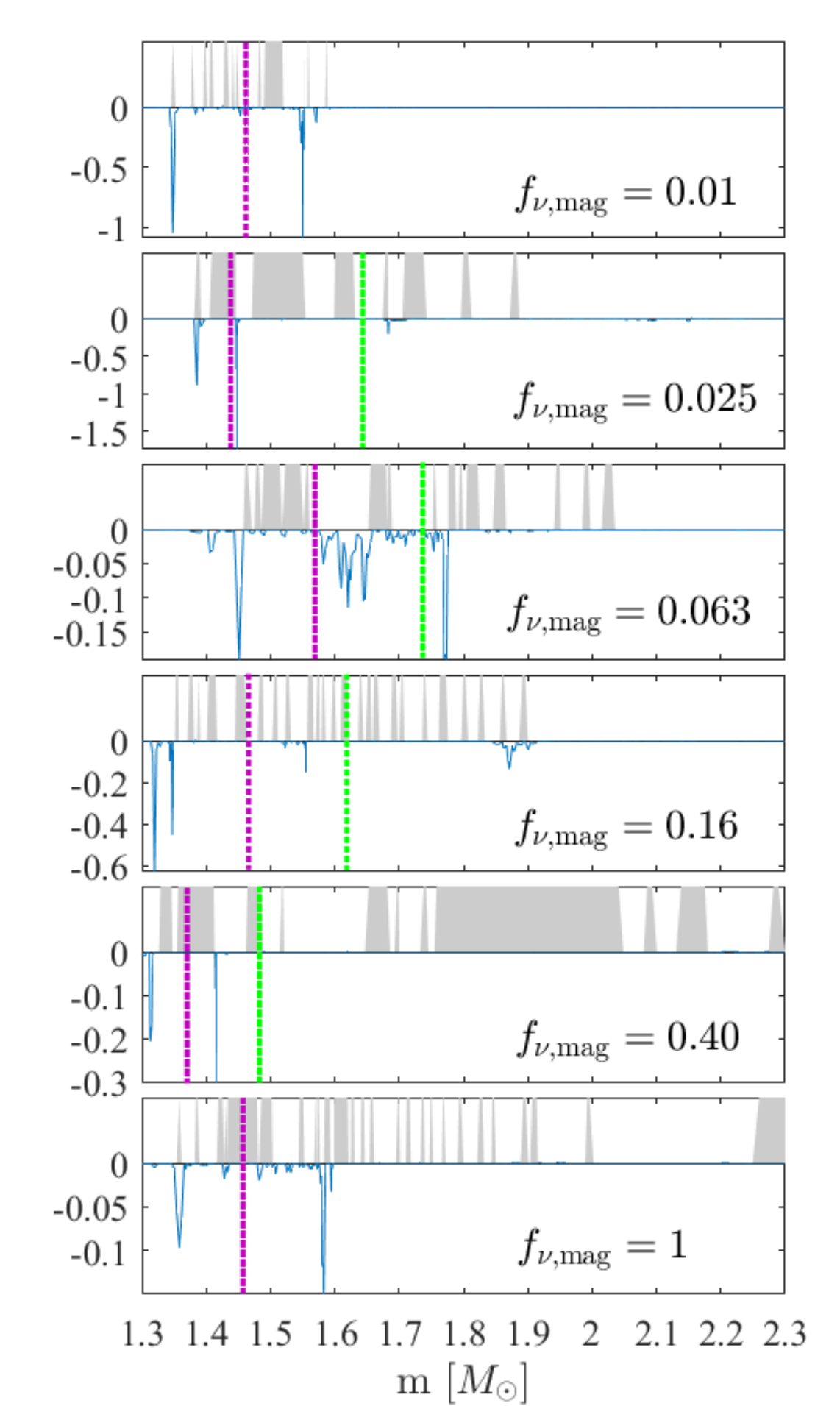} \\
\caption{Like Fig. \ref{fig:SM15eta10} but featuring models with $M_\zams=15M_\odot$, $\Omega=0.2$, $\eta=1$, and 6 equally logarithmically spaced values from 0.01 to 1 of magnetic viscosity reduction $f_{\nu,\rm{mag}}$. In the $f_{\nu,\rm{mag}}=0.01$ and $f_{\nu,\rm{mag}}=1$ cases the edge of the Si shell is unmarked as silicon and oxygen are heavily mixed above the Fe core. }
      \label{fig:SM15varf}
     \end{figure}

\subsection{Possible implications of rotational shear}
\label{subsec:shearimplicaitons}

In all the cases that we present in Figs. \ref{fig:SM15eta10} to \ref{fig:SM15varf}, zones with strong shear are seen above, and not far from, the mass coordinate of about $1.3 M_\odot$. We consider this baryonic mass because it corresponds to a final gravitational mass of the NS of about $1.2 M_\odot$, which is about the minimal value of observed NS. When the central object, the proto-NS, reaches a mass of about $0.5 M_\odot$ \citep{Janka2012} the pressure in the centre is high enough to overcome the ram pressure of the in-falling material and a shock wave starts to propagate out. This is the core bounce. The shock propagates to about $150 \km$ where it stalls (e.g., \citealt{Muller2016}). The presence of zones with strong shear above a mass of about $1.3 M_\odot$ has some implications on the supernova explosion process. 

There are many studies on the collapse of rotating cores (see section \ref{sec:intro}), and some studies that concentrate on the shear immediately after collapse. \cite{SawaiYamada2014} and \cite{Mostaetal2015}, for example, studied the collapse of a rapidly rotating pre-collapse core within the frame of neutrino explosion mechanism, e.g., they were looking for the revival of the stalled shock. We differ from most previous studies in (1) looking also at slowly rotating pre-collapse core, and (2) working in the frame of the jet feedback mechanism, and in particular the jittering-jets mechanism. 

It is hard to estimate the shear that is expected to form around the proto-NS, at a radius of about $20-30 \km$, as a result of the shear in the pre-collapse core. The reason is that after the falling material is shocked, at about $150 \km$, it enters a region of turbulence and instabilities. Nonetheless, our results point out that the shear around the proto-NS might be non-negligible in influencing instabilities and in amplifying magnetic fields, even in the slowly rotating pre-collapse cores that we study.   

\cite{Mostaetal2015} take for their magnetohydrodynamic simulations of collapsing rapidly rotating cores a rotation profile from \cite{TakiwakiKotake2011}, that in the equatorial plane is given by 
\begin{equation}
\omega(r)=2.8 \left[ 1+ \left( \frac{r}{500 \km} \right)^{2} \right]^{-1}
{\rm rad} \s^{-1}.  
\label{eq:mosta}
\end{equation}
The typical shear for this profile at a radius of $1650 \km$ and a mass of $1.5 M_\odot$ is $q_{\rm Kep} \simeq - 0.06$. This is the magnitude of the shear they obtain for specific angular momentum of $6.4\times 10^{15} \cm^2 \s^{-1}$ at $r=1650 \km$ in the pre-collapse core, which is a specific angular momentum about an order of magnitude above the values of specific angular momentum we have here at this radius (section \ref{sec:rotation}). 
  Namely, for a rotation velocity that is about an order of magnitude slower than that used by \cite{Mostaetal2015}, we find zones with shear magnitudes that are about equal to, and even larger than, those in their rotational profile. The results of \cite{Mostaetal2015} might suggest then, that the shear around the proto-NS can amplify the magnetic fields even in cases of slowly rotating pre-collapse cores. To study this effect, the results of \cite{Mostaetal2015} indicate also that one has to use very high resolution 3D magnetohydrodynamical numerical codes to simulate the effects of magnetic fields in collapsing cores (beyond our present capabilities). 
  
 There is another effect that deserves further study in futures works. The stochastic nature of the convection zones in the collapsing core might lead to the formation of an accretion belt, or even an accretion disk, around the pre-NS \citep{GilkisSoker2014}. The combined operation of this stochastic angular momentum, the SASI, the shear, and the  rotation (even if slow), might lead to amplification of instabilities and magnetic fields in a flow with preferred direction (the angular momentum direction). The basic assumption of the jittering-jets mechanism is that these lead to the launching of intermittent-stochastic jets. The new addition here is that the zones with large shear might facilitate the formation of intermittent accretion disks/belts, and hence the launching of jets. 

  Many of the high-shear zones are adjacent to convective layers. This has two implications. Shear and adjacent convection lead to amplification of magnetic fields before the collapse, through the operation of a dynamo. After the collapse, the accreted gas from these zones will possess both stochastic variation of the specific angular momentum \citep{GilkisSoker2014} and sheared inflow. These are necessary for the jittering-jets mechanism.  
  
\subsection{Cases of black hole formation}
\label{subsec:blackholes}
  
 In case the star does not explode after the formation of a NS, material continues to be accreted and forms a BH. For that possibility, in Figs. \ref{fig:SM15VarEta1t10} and \ref{fig:SM30varEta1t10} we present the shear in the outer layers of the core of some cases. We note the following properties. (1) There is a strong shear just above the outer boundary of the oxygen layer. (2) There is a desert of high-shear zones in the range from about $2.5 M_\odot$ to the outer boundary of the oxygen layer. The only such zone is at a mass coordinate of $2.7 M_\odot$ (upper panel of Fig. \ref{fig:SM30varEta1t10}), but it is far from a convective zone. 
\begin{figure}
      \vskip 0.5 cm
    \hskip -0.5 cm 
  \includegraphics*[scale=0.63]{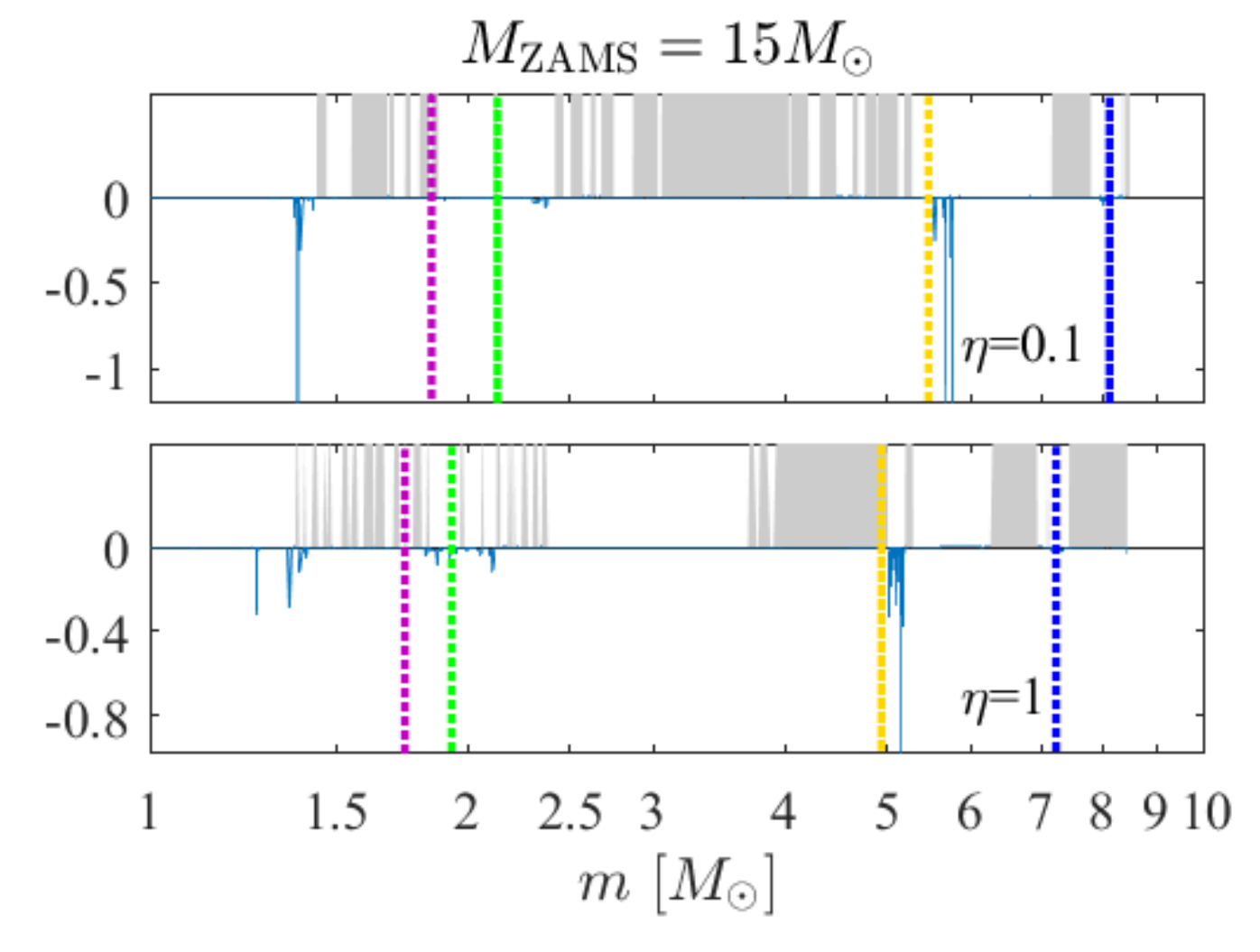} \\
\caption{Shear and convection in the $1-10M_\odot$ range of the mass coordinate for models of $M_\zams=15 M_\odot$,  $\Omega=0.5$, and for two different mass-loss scaling factors $\eta$. Pink lines mark the edge of the Fe core, green lines mark the edge of the Si shell, yellow lines mark the edge of the O shell and blue lines mark the edge of the He shell. Shears that dropped lower than $-1$ were cut at this value in the graphs presented here. The model with $\eta=0.1$ had a minimum shear of $-11$ in that range.}
      \label{fig:SM15VarEta1t10}
\end{figure}
\begin{figure}
      \vskip 0.5 cm
    \hskip -0.5 cm
  \includegraphics*[scale=0.64]{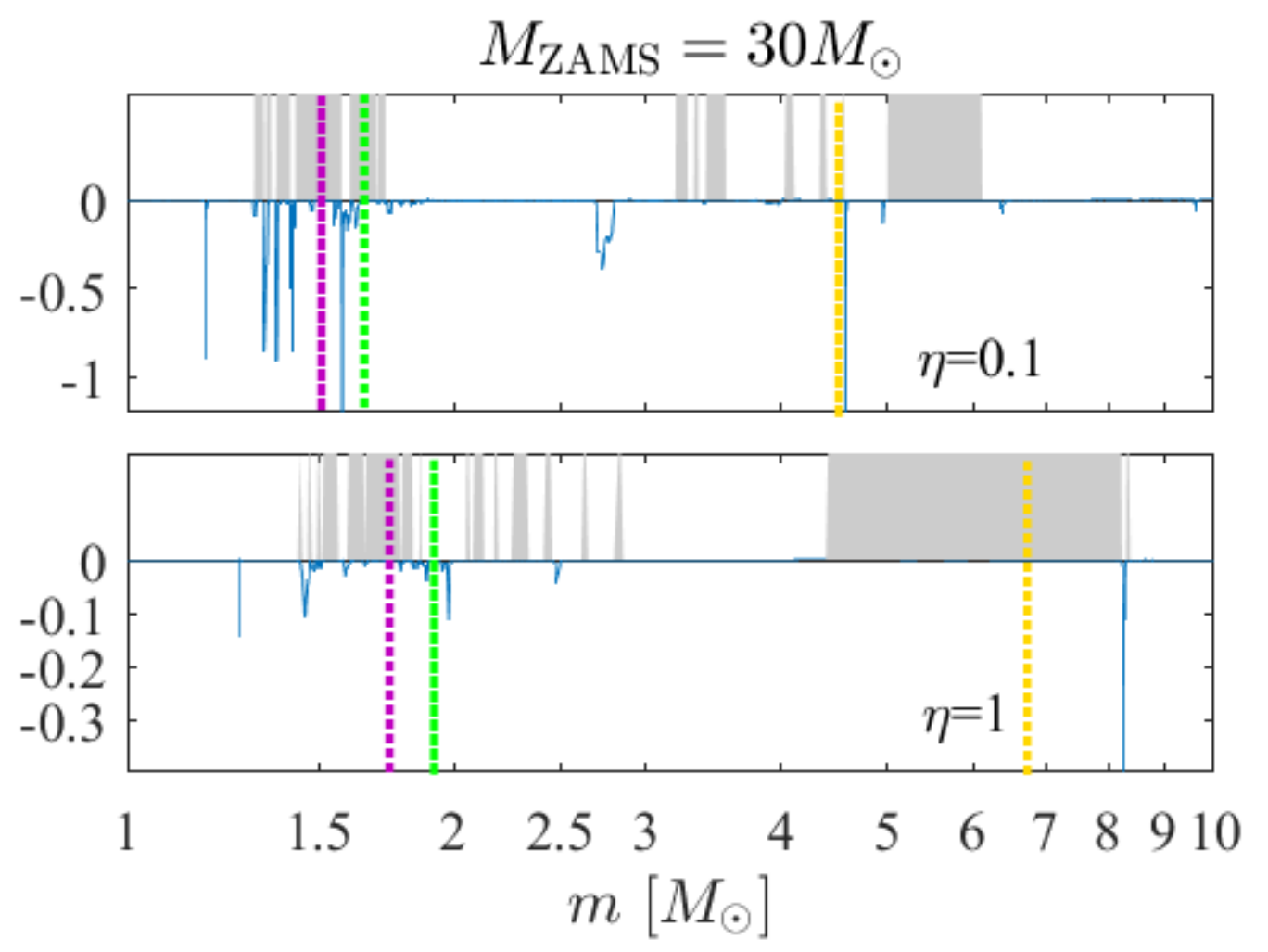} \\
\caption{Like Fig. \ref{fig:SM15VarEta1t10}, but for models with $M_\zams=30 M_\odot$. 
The model with $\eta=0.1$ had a minimum shear of $-4$ that we cut at $-1$.}
      \label{fig:SM30varEta1t10}
\end{figure}

\subsection{Main results related to rotational shear}
\label{subsec:shearmain}
We can summarize the main results discussed in this section as follows. Several high-shear zones, some that are adjacent to convective zones, exist in the mass range $\approx 1.3-2.2 M_\odot$ near and outside the boundary of the iron core just before collapse. These high-shear zones adjacent to convective zones are likely to amplify magnetic fields before the collapse. According to the assumption of the jittering-jets explosion mechanism, such zones are likely to lead to the formation of stochastic-intermittent accretion belts/disks around the newly-born NS. These accretion belts/disks  launch jets, that according to the jet feedback mechanism, explode the star. There is a desert of such zones in the mass range $\approx 2.5-4.5 M_\odot$. Further such zones appear outside the oxygen layer, and will play a role in the case that the inner core does not bring to explosion, and a BH forms at the centre.  

\section{SUMMARY}
\label{sec:summary}

In performing this study we were motivated by the failure of 3D numerical simulations to consistently and persistently explode massive stars by the delayed neutrino mechanism. One possible alternative is to explode massive stars with jets (see discussion in section \ref{sec:intro}). Jets are launched by accretion disks or accretion belts around the newly-born NS or BH, with the amplification of magnetic fields in these disks/belts. The rotation profile and the rotational shear are important for the magnetic field amplification (dynamo).  We set the goal to study the influence of mass-loss rate and coupling (magnetic viscosity) on the rotation profile and the rotational shear in the core just before core-collapse, for models with different initial masses and rotation velocities.

We evolved a set of single massive rotating stars within a large range of parameters (18 values of initial rotation velocity, 10 values of mass-loss rate, 11 values of magnetic viscosity and 2 values of initial mass). We found, as described in section \ref{sec:rotation} and shown in Figs. \ref{fig:jm15}-\ref{fig:jcore30}, that the pre-collapse specific angular momentum profiles of the inner layers of the core, from iron to oxygen, of our stellar models do not reach the values required to form a fully developed Keplerian accretion disk around the newly-born NS or BH. Our calculations do not consider the formation of intermittent-stochastic accretion disks/belts that are formed by instabilities before and after collapse.

Despite the relatively slow pre-collapse core rotation in the majority of our stellar models (see section \ref{sec:rotation}), we found, as shown in Figs. \ref{fig:SM15eta10}-\ref{fig:SM15varf}, that there are zones of large rotational shear above the iron core, and that many of them are adjacent to convective zones. These regions are located in a mass coordinate that is appropriate for the formation of a NS with a gravitational mass of $\approx 1.2-2 M_\odot$. In case the core does not explode after the inner $\approx 2.5 M_\odot$ collapse, we find that the next region of high rotational shear zones is above the oxygen shell at a mass coordinate of $\ga 4.5 M_\odot$, as shown in Figs. \ref{fig:SM15VarEta1t10} and \ref{fig:SM30varEta1t10}. This corresponds to the formation of a BH.  

These results hint that even slowly rotating pre-collapse cores might amplify magnetic fields in the pre-collapse core, and in the region above the newly-formed NS after collapse. Such an amplification is required in the jet feedback explosion mechanism, and in particular it is relevant to the jittering jets model (see section \ref{sec:intro}). 

Our results suggest that in simulating the explosion of the cores of massive stars one has to consider rotational profiles that include zones of high rotational shear as we find in the present study. Analytical smoothed fitting to the rotation profile might miss essential processes. Such high-shear zones might play a significant role in the stochastic/unstable/turbulent flow beyond the stalled shock, where gas is accreted on to the newly-born NS or BH.  

\section*{Acknowledgments}
{{{ We thank the referee, Georges Meynet, for his very detailed and valuable comments. }}}
This research was supported by the Israel Science Foundation and by the Pazy Research Foundation.
N.Z. thanks the support of the PEF's Alfred and Anna Grey Excellence Scholarship.

\appendix
\section*{APPENDIX: Pre-collapse properties of the stellar models}
\label{app:models}

Table \ref{tab:M15} presents several properties of interest for all our models with $M_\zams=15M_\odot$, and Table \ref{tab:M30} presents the models with $M_\zams=30M_\odot$. The first three columns list the initial conditions and model assumptions, with the rest showing the resulting pre-collapse properties. We list parameters of compact objects formed through three scenarios: ($i$) The inner core explodes the star, and a NS is formed, with a mass approximately that of the iron core. ($ii$) A BH forms from the CO core, with the rest being expelled. ($iii$) All material below the hydrogen envelope forms a BH, unless is has sufficient angular momentum to avoid accretion.
\begin{table*}
\centering
\caption{Model parameters for $M_\zams=15 M_\odot$}
\begin{threeparttable}
\begin{tabular}{ccccccccccccccccc}
\hline
$\Omega$ & $\eta$ & $f_{\nu, \rm{mag}}$ & $M_\mathrm{f}$ & $R_\mathrm{ph}$ & $L_\mathrm{ph}$ & $T_\mathrm{eff}$ & $M_\mathrm{Fe}$ & $J_\mathrm{Fe}$ & $P_\mathrm{NS}$ & $M_\mathrm{CO}$ & $J_\mathrm{CO}$ & $a_\mathrm{CO}$ & $M_\mathrm{BH}$ & $J_\mathrm{BH}$ & $a_\mathrm{BH}$ & class\\
\hline
0.05& 0.1 & 1 & 14.65 & 1117 & 1.08 & 3132 & 1.57 & 0.02 & 626 & 3.39 & 0.10 & 0.001 & 5.15 & 0.25 & 0.001 & \textcolor{Red}{RSG}\\
0.1 &	0.1 &	1 &	14.65 &	1087 &	1.02 &	3133 &	1.51 &	0.06 &	146 &	3.40 &	0.43 &	0.004 &	5.17 &	1.30 &	0.006 & \textcolor{Red}{RSG}\\
0.15 &	0.1 &	1 &	14.69 &	1083 &	1.03 &	3142 &	1.56 &	0.14 &	68 &	3.34 &	1.10 &	0.011 &	5.10 &	3.46 &	0.015 & \textcolor{Red}{RSG}\\
0.2 &	0.1 &	1 &	14.66 &	1090 &	1.01 &	3121 &	1.45 &	0.22 &	39 &	3.39 &	2.30 &	0.023 &	5.15 &	7.28 &	0.031 & \textcolor{Red}{RSG}\\
0.25 &	0.1 &	1 &	14.66 &	1121 &	1.11 &	3150 &	1.49 &	0.19 &	46 &	3.42 &	1.91 &	0.019 &	5.18 &	5.92 &	0.025 & \textcolor{Red}{RSG}\\
0.3 &	0.1 &	1 &	14.70 &	1085 &	1.03 &	3139 &	1.51 &	0.29 &	31 &	3.27 &	2.42 &	0.026 &	5.02 &	7.80 &	0.035 & \textcolor{Red}{RSG}\\
0.35 &	0.1 &	1 &	14.23 &	1142 &	1.28 &	3234 &	1.51 &	0.02 &	466 &	3.71 &	0.17 &	0.001 &	5.54 &	0.47 &	0.002 & \textcolor{Red}{RSG}\\
0.4 &	0.1 &	1 &	14.10 &	1145 &	1.39 &	3298 &	1.56 &	0.04 &	235 &	3.76 &	0.35 &	0.003 &	5.60 &	1.00 &	0.004 & \textcolor{Red}{RSG}\\
0.45 &	0.1 &	1 &	13.14 &	1114 &	1.56 &	3440 &	1.53 &	0.29 &	32 &	4.12 &	4.43 &	0.030 &	6.07 &	12.97 &	0.040 & \textcolor{Red}{RSG}\\
0.5 &	0.1 &	1 &	8.48 &	127 &	1.87 &	10677 &	1.81 &	1.09 &	11 &	5.64 &	21.30 &	0.076 &	8.04 &	59.26 &	0.104 & \textcolor{Blue}{BSG}\\
0.55 &	0.1 &	1 &	9.67 &	6.1 &	1.92 &	48910 &	1.80 &	2.39 &	5 &	6.99 &	67.40 &	0.157 &	9.34 &	162.63 &	0.212 & \textcolor{Magenta}{WR}\\
0.6 &	0.1 &	1 &	9.07 &	1.6 &	2.69 &	103368 &	1.61 &	3.76 &	3 &	5.05 &	40.90 &	0.182 &	8.03 &	268.53 &	0.473 & \textcolor{Magenta}{WR}\\
0.65 &	0.1 &	1 &	10.51 &	1.7 &	2.90 &	101884 &	1.83 &	2.49 &	5 &	5.75 &	26.40 &	0.091 &	10.15 &	273.18 &	0.301 & \textcolor{Magenta}{WR}\\
\hline
0.05 &	1 &	1 &	12.42 &	1084 &	1.00 &	3122 &	1.52 &	0.69 &	13 &	3.11 &	6.01 &	0.071 &	4.84 &	20.15 &	0.098 & \textcolor{Red}{RSG}\\
0.1 &	1 &	1 &	12.18 &	1108 &	1.14 &	3189 &	1.51 &	0.54 &	17 &	3.17 &	5.45 &	0.062 &	4.90 &	17.42 &	0.082 & \textcolor{Red}{RSG}\\
0.15 &	1 &	1 &	11.44 &	1096 &	1.15 &	3210 &	1.42 &	0.51 &	17 &	3.33 &	7.85 &	0.080 &	5.11 &	27.58 &	0.120 & \textcolor{Red}{RSG}\\
0.2 &	1 &	1 &	11.62 &	1097 &	1.09 &	3168 &	1.46 &	0.36 &	24 &	3.30 &	4.05 &	0.042 &	5.07 &	12.70 &	0.056 & \textcolor{Red}{RSG}\\
0.25 &	1 &	1 &	12.00 &	1109 &	1.07 &	3136 &	1.48 &	0.27 &	33 &	3.17 &	2.56 &	0.029 &	4.92 &	8.30 &	0.039 & \textcolor{Red}{RSG}\\
0.3 &	1 &	1 &	11.33 &	1101 &	1.18 &	3228 &	1.50 &	0.58 &	16 &	3.39 &	6.69 &	0.066 &	5.17 &	20.48 &	0.087 & \textcolor{Red}{RSG}\\
0.35 &	1 &	1 &	10.59 &	1111 &	1.25 &	3256 &	1.40 &	0.24 &	34 &	3.57 &	3.90 &	0.035 &	5.40 &	12.01 &	0.047 & \textcolor{Red}{RSG}\\
0.4 &	1 &	1 &	10.05 &	1051 &	1.32 &	3397 &	1.51 &	0.86 &	11 &	3.72 &	10.74 &	0.088 &	5.58 &	34.40 &	0.126 & \textcolor{Red}{RSG}\\
0.45 &	1 &	1 &	8.12 &	880 &	1.47 &	3811 &	1.64 &	0.92 &	11 &	4.19 &	13.27 &	0.086 &	6.16 &	41.09 &	0.123 & \textcolor{Red}{RSG}\\
0.5 &	1 &	1 &	8.44 &	628 &	2.04 &	4895 &	1.73 &	1.12 &	10 &	5.07 &	20.68 &	0.091 &	7.29 &	56.88 &	0.122 & \textcolor{YellowOrange}{YSG}\\
0.55 &	1 &	1 &	8.39 &	475 &	2.00 &	5605 &	1.61 &	1.20 &	8 &	5.42 &	22.13 &	0.086 &	7.68 &	60.87 &	0.117 & \textcolor{YellowOrange}{YSG}\\
0.6 &	1 &	1 &	7.28 &	1.2 &	1.94 &	108630 &	1.54 &	0.23 &	41 &	5.21 &	4.32 &	0.018 &	7.28 &	8.38 &	0.018 & \textcolor{Magenta}{WR}\\
0.65 &	1 &	1 &	6.49 &	1.4 &	1.62 &	97181 &	1.48 &	0.58 &	15 &	4.56 &	9.84 &	0.054 &	6.49 &	32.70 &	0.088 & \textcolor{Magenta}{WR}\\
0.7 &	1 &	1 &	6.62 &	1.2 &	1.69 &	105140 &	1.47 &	0.76 &	12 &	4.64 &	12.67 &	0.067 &	6.61 &	61.46 &	0.159 & \textcolor{Magenta}{WR}\\
0.75 &	1 &	1 &	5.83 &	1.2 &	1.20 &	96964 &	1.43 &	1.07 &	8 &	4.07 &	18.89 &	0.129 &	5.55 &	95.83 &	0.354 & \textcolor{Magenta}{WR}\\
0.8 &	1 &	1 &	6.10 &	1.3 &	1.43 &	98003 &	1.50 &	0.70 &	13 &	4.32 &	12.25 &	0.075 &	6.09 &	62.05 &	0.190 & \textcolor{Magenta}{WR}\\
0.85 &	1 &	1 &	5.61 &	1.5 &	1.37 &	90161 &	1.46 &	0.99 &	9 &	3.93 &	16.43 &	0.121 &	5.25 &	85.96 &	0.354 & \textcolor{Magenta}{WR}\\
0.9 &	1 &	1 &	5.68 &	1.2 &	1.26 &	96779 &	1.46 &	0.86 &	10 &	3.96 &	12.89 &	0.093 &	5.55 &	83.44 &	0.307 & \textcolor{Magenta}{WR}\\
\hline
0.2 &	0.2 &	1 &	14.30 &	1106 &	1.21 &	3235 &	1.52 &	0.24 &	39 &	3.38 &	2.24 &	0.022 &	5.14 &	6.95 &	0.030 & \textcolor{Red}{RSG}\\
0.2 &	0.3 &	1 &	14.12 &	1128 &	1.05 &	3097 &	1.33 &	0.09 &	88 &	3.26 &	1.00 &	0.011 &	5.01 &	3.19 &	0.014 & \textcolor{Red}{RSG}\\
0.2 &	0.4 &	1 &	13.73 &	1119 &	1.07 &	3126 &	1.43 &	0.21 &	39 &	3.31 &	2.24 &	0.023 &	5.06 &	7.07 &	0.031 & \textcolor{Red}{RSG}\\
0.2 &	0.5 &	1 &	13.61 &	1116 &	1.05 &	3115 &	1.56 &	0.28 &	35 &	3.20 &	2.21 &	0.025 &	4.93 &	7.16 &	0.033 & \textcolor{Red}{RSG}\\
0.2 &	0.6 &	1 &	12.93 &	1099 &	1.04 &	3131 &	1.57 &	0.53 &	18 &	3.33 &	4.88 &	0.050 &	5.11 &	15.75 &	0.068 & \textcolor{Red}{RSG}\\
0.2 &	0.7 &	1 &	13.27 &	1088 &	1.03 &	3136 &	1.52 &	0.34 &	27 &	3.10 &	2.85 &	0.034 &	4.84 &	9.53 &	0.046 & \textcolor{Red}{RSG}\\
0.2 &	0.8 &	1 &	12.75 &	1108 &	1.10 &	3164 &	1.46 &	0.27 &	32 &	3.18 &	3.01 &	0.034 &	4.91 &	9.78 &	0.046 & \textcolor{Red}{RSG}\\
0.2 &	0.9 &	1 &	11.50 &	1107 &	1.17 &	3211 &	1.49 &	0.25 &	36 &	3.41 &	2.39 &	0.023 &	5.20 &	7.72 &	0.032 & \textcolor{Red}{RSG}\\
\hline
0.2 &	1 &	0.01 &	11.35 &	1095 &	1.28 &	3303 &	1.46 &	2.20 &	4 &	3.40 &	23.81 &	0.234 &	5.19 &	75.65 &	0.319 & \textcolor{Red}{RSG}\\
0.2 &	1 &	0.0158 &	11.35 &	1102 &	1.13 &	3187 &	1.53 &	2.26 &	4 &	3.38 &	21.06 &	0.209 &	5.16 &	67.41 &	0.287 & \textcolor{Red}{RSG}\\
0.2 &	1 &	0.0251 &	12.00 &	1097 &	1.08 &	3164 &	1.45 &	1.50 &	6 &	3.24 &	16.02 &	0.173 &	4.98 &	52.71 &	0.241 & \textcolor{Red}{RSG}\\
0.2 &	1 &	0.0398 &	11.37 &	1109 &	1.18 &	3213 &	1.57 &	1.87 &	5 &	3.37 &	16.56 &	0.165 &	5.16 &	53.98 &	0.230 & \textcolor{Red}{RSG}\\
0.2 &	1 &	0.0631 &	11.96 &	1098 &	1.10 &	3174 &	1.58 &	1.23 &	8 &	3.25 &	9.95 &	0.107 &	4.99 &	31.97 &	0.146 & \textcolor{Red}{RSG}\\
0.2 &	1 &	0.1 &	11.38 &	1108 &	1.22 &	3244 &	1.46 &	1.26 &	7 &	3.37 &	13.32 &	0.133 &	5.16 &	44.60 &	0.191 & \textcolor{Red}{RSG}\\
0.2 &	1 &	0.1585 &	11.70 &	1105 &	1.15 &	3196 &	1.47 &	0.83 &	11 &	3.28 &	10.92 &	0.115 &	5.04 &	36.80 &	0.164 & \textcolor{Red}{RSG}\\
0.2 &	1 &	0.2512 &	11.66 &	1107 &	1.12 &	3175 &	1.47 &	0.79 &	11 &	3.30 &	9.18 &	0.096 &	5.06 &	30.09 &	0.133 & \textcolor{Red}{RSG}\\
0.2 &	1 &	0.3981 &	11.92 &	1096 &	1.08 &	3161 &	1.37 &	0.28 &	28 &	3.24 &	2.91 &	0.032 &	4.99 &	9.30 &	0.042 & \textcolor{Red}{RSG}\\
0.2 &	1 &	0.631 &	11.73 &	1113 &	1.10 &	3156 &	1.42 &	0.25 &	33 &	3.27 &	3.48 &	0.037 &	5.03 &	10.99 &	0.049 & \textcolor{Red}{RSG}\\
\hline
\end{tabular}
\footnotesize
\begin{tablenotes}
Note: $\Omega \equiv \left(\omega / \omega_\mathrm{crit}\right)_\zams$, with $\omega_\mathrm{crit}$ according to equation (\ref{eq:wcrit}). $\eta$ is the mass-loss scaling factor. $f_{\nu, \rm{mag}}$ is the scaling factor for the magnetic viscosity. $M_\mathrm{f}$ is the pre-collapse stellar mass in $M_\odot$. $R_\mathrm{ph}$ is the photosphere radius in $R_\odot$. $L$ is the luminosity in $10^5 \times L_\odot$. $T_\mathrm{eff}$ is the effective temperature in $K$. $M_\mathrm{Fe}$ is the pre-collapse iron core mass in $M_\odot$, and $J_\mathrm{Fe}$ is its total angular momentum in $10^{48} \g \cm^2 \s^{-1}$. $P_\mathrm{NS}$ is the rotation period in $\mathrm{ms}$ of a NS with $M_\mathrm{Fe}$ and $J_\mathrm{Fe}$, calculated using $I=1.3\times 10^{45} \left(M_\mathrm{Fe} / 1.4 M_\odot \right)^{3/2}$ for the NS moment of inertia \citep{LattimerSchutz2005,Metzger2015}. $M_\mathrm{CO}$ is the mass of the pre-collapse carbon-oxygen core in $M_\odot$, and $J_\mathrm{CO}$ is its total angular momentum in $10^{48} \g \cm^2 \s^{-1}$. $a_\mathrm{CO}=cJ_\mathrm{CO}/GM_\mathrm{CO}^2$ is the spin parameter of a Kerr BH if it is formed from the carbon-oxygen core. $M_\mathrm{BH}$ is the mass up to the outer edge of the helium layer, or up to the point where $j/j_\mathrm{LSO,Sch}>1$, whichever comes first. We consider this the maximal possible BH mass. $J_\mathrm{BH}$ is the corresponding total angular momentum in $10^{48} \g \cm^2 \s^{-1}$, and $a_\mathrm{BH}=cJ_\mathrm{BH}/GM_\mathrm{BH}^2$. The final column indicates our classification for the stellar type for each pre-collapse model, as described in section \ref{sec:rotation}.
\end{tablenotes}
\end{threeparttable}
\label{tab:M15}
\end{table*}

\begin{table*}
\centering
\caption{Model parameters for $M_\zams=30 M_\odot$}
\begin{threeparttable}
\begin{tabular}{ccccccccccccccccc}
\hline
$\Omega$ & $\eta$ & $f_{\nu, \rm{mag}}$ & $M_\mathrm{f}$ & $R_\mathrm{ph}$ & $L_\mathrm{ph}$ & $T_\mathrm{eff}$ & $M_\mathrm{Fe}$ & $J_\mathrm{Fe}$ & $P_\mathrm{NS}$ & $M_\mathrm{CO}$ & $J_\mathrm{CO}$ & $a_\mathrm{CO}$ & $M_\mathrm{BH}$ & $J_\mathrm{BH}$ & $a_\mathrm{BH}$ & class\\
\hline
0.05 &	1 &	1 &	12.75 &	4.4 &	4.07 &	69286 &	1.48 &	1.19 &	7 &	3.82 &	7.00 &	0.054 &	12.58 &	163.12 &	0.117 & \textcolor{Magenta}{WR}\\
0.1 &	1 &	1 &	12.94 &	387 &	4.59 &	7641 &	1.82 &	1.11 &	11 &	6.83 &	13.91 &	0.034 &	12.00 &	91.82 &	0.072 & \textcolor{Blue}{BSG}\\
0.15 &	1 &	1 &	12.62 &	4.7 &	4.34 &	68354 &	1.85 &	0.54 &	23 &	8.01 &	10.81 &	0.019 &	12.35 &	45.04 &	0.034 & \textcolor{Magenta}{WR}\\
0.2 &	1 &	1 &	12.66 &	130 &	4.09 &	12813 &	1.77 &	0.29 &	40 &	5.51 &	2.06 &	0.008 &	11.95 &	24.93 &	0.020 & \textcolor{Blue}{BSG}\\
0.25 &	1 &	1 &	13.01 &	146 &	4.42 &	12346 &	1.70 &	0.34 &	32 &	8.01 &	9.03 &	0.016 &	11.98 &	28.96 &	0.023 & \textcolor{Blue}{BSG}\\
0.3 &	1 &	1 &	12.55 &	111 &	4.22 &	14003 &	1.90 &	0.40 &	33 &	9.23 &	13.91 &	0.019 &	12.16 &	31.28 &	0.024 & \textcolor{Blue}{BSG}\\
0.35 &	1 &	1 &	12.46 &	2.3 &	3.03 &	88967 &	1.48 &	1.05 &	8 &	2.22 &	1.85 &	0.043 &	12.46 &	103.57 &	0.076 & \textcolor{Magenta}{WR}\\
0.4 &	1 &	1 &	14.40 &	305 &	4.43 &	8539 &	1.69 &	0.47 &	23 &	9.32 &	17.77 &	0.023 &	12.51 &	40.98 &	0.030 & \textcolor{Blue}{BSG}\\
0.45 &	1 &	1 &	12.25 &	0.9 &	3.87 &	149070 &	1.56 &	1.23 &	8 &	7.92 &	23.34 &	0.042 &	12.25 &	94.96 &	0.072 & \textcolor{Magenta}{WR}\\
0.5 &	1 &	1 &	12.22 &	0.7 &	3.91 &	176329 &	1.73 &	0.71 &	16 &	8.26 &	19.76 &	0.033 &	12.22 &	51.38 &	0.039 & \textcolor{Magenta}{WR}\\
0.55 &	1 &	1 &	13.54 &	1.2 &	4.29 &	137783 &	1.54 &	0.35 &	27 &	8.55 &	9.22 &	0.014 &	13.54 &	28.58 &	0.018 & \textcolor{Magenta}{WR}\\
0.6 &	1 &	1 &	10.54 &	0.7 &	2.94 &	156029 &	1.62 &	0.39 &	26 &	7.26 &	7.87 &	0.017 &	10.54 &	25.87 &	0.026 & \textcolor{Magenta}{WR}\\
0.65 &	1 &	1 &	8.95 &	0.9 &	2.29 &	132055 &	1.56 &	0.45 &	21 &	2.52 &	0.77 &	0.014 &	8.95 &	24.19 &	0.034 & \textcolor{Magenta}{WR}\\
0.7 &	1 &	1 &	8.37 &	0.8 &	2.18 &	139669 &	1.51 &	0.08 &	109 &	4.20 &	0.40 &	0.003 &	8.37 &	3.16 &	0.005 & \textcolor{Magenta}{WR}\\
0.75 &	1 &	1 &	7.87 &	0.8 &	2.06 &	136999 &	1.60 &	0.05 &	194 &	5.48 &	0.77 &	0.003 &	7.87 &	1.75 &	0.003 & \textcolor{Magenta}{WR}\\
0.8 &	1 &	1 &	7.84 &	0.8 &	2.05 &	141948 &	1.69 &	0.35 &	31 &	5.49 &	5.47 &	0.021 &	7.84 &	12.38 &	0.023 & \textcolor{Magenta}{WR}\\
0.85 &	1 &	1 &	7.79 &	0.7 &	2.22 &	149138 &	1.60 &	0.38 &	27 &	5.45 &	6.30 &	0.024 &	7.79 &	14.47 &	0.027 & \textcolor{Magenta}{WR}\\
\hline
0.05 &	0.1 &	1 &	27.95 &	784 &	3.38 &	4976 &	1.92 &	1.47 &	9 &	7.30 &	23.45 &	0.050 &	12.36 &	130.12 &	0.097 & \textcolor{YellowOrange}{YSG}\\
0.1 &	0.1 &	1 &	27.80 &	923 &	3.87 &	4742 &	1.63 &	0.46 &	22 &	7.65 &	11.54 &	0.022 &	11.39 &	34.58 &	0.030 & \textcolor{YellowOrange}{YSG}\\
0.15 &	0.1 &	1 &	27.23 &	857 &	3.11 &	4658 &	1.78 &	0.17 &	71 &	8.89 &	5.34 &	0.008 &	11.77 &	11.80 &	0.010 & \textcolor{YellowOrange}{YSG}\\
0.2 &	0.1 &	1 &	26.20 &	838 &	2.21 &	4328 &	1.65 &	0.16 &	67 &	4.20 &	0.68 &	0.004 &	11.71 &	10.38 &	0.009 & \textcolor{YellowOrange}{YSG}\\
0.25 &	0.1 &	1 &	25.71 &	812 &	2.19 &	4385 &	1.62 &	0.60 &	17 &	3.80 &	2.09 &	0.017 &	12.46 &	48.18 &	0.035 & \textcolor{YellowOrange}{YSG}\\
0.3 &	0.1 &	1 &	26.49 &	1055 &	2.75 &	4072 &	1.67 &	0.82 &	13 &	5.10 &	6.06 &	0.026 &	11.69 &	61.65 &	0.051 & \textcolor{YellowOrange}{YSG}\\
0.35 &	0.1 &	1 &	23.81 &	597 &	3.44 &	5727 &	1.66 &	0.38 &	28 &	4.75 &	2.10 &	0.011 &	13.91 &	37.63 &	0.022 & \textcolor{YellowOrange}{YSG}\\
0.4 &	0.1 &	1 &	23.89 &	470 &	1.72 &	5421 &	1.37 &	0.73 &	11 &	2.69 &	3.00 &	0.047 &	15.04 &	130.82 &	0.066 & \textcolor{YellowOrange}{YSG}\\
0.45 &	0.1 &	1 &	19.64 &	319 &	7.54 &	9535 &	1.47 &	0.35 &	25 &	2.83 &	1.25 &	0.018 &	18.02 &	75.64 &	0.026 & \textcolor{Blue}{BSG}\\
0.5 &	0.1 &	1 &	19.82 &	2.0 &	6.90 &	118952 &	1.51 &	4.66 &	2 &	4.58 &	33.26 &	0.180 &	19.43 &	1521.18 &	0.458 & \textcolor{Magenta}{WR}\\
\hline
\end{tabular}
\footnotesize
\begin{tablenotes}
Note: Same as table \ref{tab:M15}, but for $M_\zams=30M_\odot$.
\end{tablenotes}
\end{threeparttable}
\label{tab:M30}
\end{table*}

\ifmnras
	\bibliographystyle{mnras}
	
\label{lastpage}
\end{document}